\documentclass[aps,prl,twocolumn,superscriptaddress,showpacs]{revtex4}
\usepackage{graphicx}
\usepackage{dcolumn}
\usepackage{color}
\usepackage[breaklinks,colorlinks = true,linkcolor = blue,
urlcolor=blue,citecolor=blue]{hyperref}

\begin{document}
\title{Dynamical quantum phase transitions}

\author{A.A. Zvyagin}
\affiliation{Max-Planck Institut f\"ur Physik komplexer Systeme, Noethnitzer
Str., 38, D-01187, Dresden, Germany}
\affiliation{B.I.~Verkin Institute for Low Temperature Physics and
Engineering of the National Academy of Sciences of Ukraine,
Nauky Ave., 47, Kharkov, 61103, Ukraine}

\begin{abstract}
During recent years the interest to dynamics of quantum systems has grown considerably. Quantum many body systems out of equilibrium often manifest behavior, different from the one predicted by standard statistical mechanics and thermodynamics in equilibrium. Since the dynamics of a many body quantum system typically involve many excited eigenstates, with a non-thermal distribution, the time evolution of such a system provides an unique way for investigation of non-equilibrium quantum statistical mechanics. Last decade such new subjects like quantum quenches, thermalization, pre-thermalization, equilibration, generalized Gibbs ensemble, etc. are among the most attractive topics of investigation in modern quantum physics. One of the most interesting themes in the study of dynamics of quantum many-body systems out of equilibrium is connected with the recently proposed important concept of dynamical quantum phase transitions. During the last few years a great progress has been achieved in studying of those singularities in the time dependence of characteristics of quantum mechanical systems, in particular, in understanding how the quantum critical points of equilibrium thermodynamics affect their dynamical properties. Dynamical quantum phase transitions reveal universality, scaling, connection to the topology, and many other interesting features. Here we review the recent achievements of this quickly developing part of low temperature quantum physics. The study of dynamical quantum phase transitions is especially important in context of their connection to the problem of the modern theory of quantum information, where namely non-equilibrium dynamics of many-body quantum system plays the major role. 
  
\end{abstract}

\date{\today}

\maketitle


\section{Introduction}

The problem of phase transitions is always interesting for physicists \cite{Max}. In the vicinity of a phase transition the main characteristics of the system are changed drastically under relatively small changes of the governing parameters (such as the temperature, external fields, pressure, etc.) \cite{Ma}. Phase transition points (lines) divide thermodynamic states of matter, phases. Each of such phases is determined by the order parameter (or its absence). Hence, the investigation of phase transitions permits us to better understand the nature of the properties of phases of matter without large changes of external conditions. On the other hand, it helps us to understand the occurrence of the discontinuities of thermodynamic functions at the phase transitions. Phase transitions are usually classified by their order. The latter is determined by the divergence of the derivative of the thermodynamic potential at the phase transition. Usually one studies the first order transitions (in which two phases can coexist at the same values of the governing parameter), and the second order ones (or the continuous phase transitions), where phases do not coexist. 

Until the last three decades physicists mostly study phase transitions occurring at nonzero temperatures. The characteristic feature of such phase transitions is that namely thermal fluctuations destroy the long range order. When describing continuous phase transitions it is often useful to consider the variable $t = |T-T_c|/T_c$, (do not confuse with time), where $T$ is the temperature, and $T_c$ is the critical value of the temperature, at which the phase transition takes place. Then the correlation length $\xi$ diverges as $\xi \propto |t|^{-\nu}$, and the correlation (or equilibration) time $\tau_c$ is divergent as $\tau_c \propto 
\xi^z \propto |t|^{-\nu z}$. Here $\nu$ is called the correlation length critical exponent, and $z$ is the dynamical exponent. It means that spatial and dynamical correlations become long-ranged at the transition point. There are no other universal space and time scales in the vicinity of the phase transition, except of $\xi$ and $\tau_c$, which become infinite at the critical point. That means that fluctuations occur at all time and space scales, and the system is scale invariant. Such a scale-invariant situation is usually described by the number of critical exponents, which determine the behavior of the specific heat ($\alpha$) order parameter ($\beta$), susceptibility ($\gamma$), critical isotherm ($\delta$), and correlation function ($\eta$) \cite{Ma}. The values of the critical exponents are connected via so-called scaling 
\begin{eqnarray}
&&2-\alpha = 2\beta +\gamma \ , \nonumber \\
&&2-\alpha = \beta(\delta +1) \ ,
\end{eqnarray}
and hyper-scaling relations
\begin{eqnarray}
&&2-\alpha = d\nu \ ,  \nonumber \\
&&\gamma =(2-\eta)\nu \ , 
\end{eqnarray}
where $d$ is the dimension of space. Since in classical statistical mechanics statics and dynamics totally decouple from each other, the dynamical exponent $z$ is independent of other critical exponents. 
Remarkable feature of the second order phase transitions is their universality.  The latter is determined by the symmetry of the order parameter in the ordered phase and by the dimension $d$. 

However, during the last decades, other phase transitions attract the attention of physicists. In those phase transitions, which take place at zero temperature (in the ground state)  the non-thermal governing parameter $g$ (like the pressure, magnetic field, chemical composition, etc.) regulates their behavior. Such phase transitions are known now as quantum phase transitions, or quantum critical points \cite{qpt}.  At the point $g=g_c$ (called the quantum critical point) quantum, not thermal, fluctuations destroy the long range ordering. The characteristic energy scale, at which quantum fluctuations are essential is $\hbar \omega_c$. Quantum fluctuations become more important for $\hbar \omega_c > T$ (we use the units, in which Boltzmann's constant is unity $k_B=1$). When the typical frequency scale goes to zero, we have $\hbar \omega_c \propto |t|^{\nu z}$. For $|t| <T_c^{1/\nu z}$ quantum fluctuations become unimportant, and one can use classical description of phase transitions. On the other hand, in the ground state the behavior of the characteristics of the phase transition become totally quantum, and we have $\hbar \omega_c \propto |g-g_c|^{\nu z}$.  

While the behavior of equilibrium phase transitions is by now relatively well understood, the behavior of physical systems, especially, of quantum systems out of equilibrium, is understood much less. Quantum systems out of equilibrium, e.g., after abrupt changes of their parameters, are basically not susceptible to general principles of equilibrium systems. This is why,  studies of non-equilibrium dynamics of quantum many-body models are necessary for the fundamental understanding of how mechanics emerges under the unitary time evolution. The time evolution of averages depends on the initial state through the values of a large number of parameters of the quantum system. It disagrees with the standard ensembles of statistical mechanics, which use few conserved values of the dynamical system and usually describe the behavior after relaxation. In the isolated system the energy is a conserved quantity. In the absence of other conserved quantities generic isolated systems are believed to relax to thermal equilibrium, i.e., to the Gibbs ensemble with an effective temperature, known as thermalization \cite{therm}. Thermalization should occur independently of the initial state of the system, and it is important to study the case in which the system is initialized in the highly excited eigenstate at the finite density of the energy. The dynamics then is simple. However the thermalization requires the statistical mechanics to be encoded in the chosed eigenstate, which is formulated as the eigenstate thermalization hypothesis \cite{therm}. The system remains in a pure state at all times, and the reduced density matrix of a small subsystem should take a Gibbs form with the effective temperature, depending only on the energy density of the chosen state. The expectation values of local observables in this eigenstate are smooth functions of the energy, which coincide with the microcanonical ensemble at the corresponding density of the energy. On the other hand, integrable systems evolve to a generalized Gibbs ensemble \cite{RDYO} due to the presence of the infinite number of integrals of motion in integrable systems. Contrary, models with weak integrability-breaking interactions exhibit transient behavior, with local observables relaxing to non-thermal values, known as pre-thermalization \cite{MK}.  Common believe is that at "sufficiently long times" pre-termalized systems thermalize. 

 Abrupt changes of some parameters lead to the unitary time evolution, and the final (long time) state strongly depends on the type of the system. Their studies can provide the information of how fast correlations spread in quantum systems, whether averages can decay to some time-independent values, and which parameters can govern those processes. The study of dynamics of the quantum coherence is very important for the modern theory of quantum computation, where namely sudden changes are used to govern the behavior of ensembles of qubits \cite{NH}. On the other hand, the study of sudden changes is very important in the context of experiments on ultracold gases, \cite{ult} THz pulses \cite{THz} observed in solids, \cite{exp} or high magnetic field experiments in pulse fields. \cite{HMF}  For ultracold gases, for instance, the coherence is maintained for much longer times than for usual condensed matter, and the time evolution of a quantum system after the abrupt changes  has become an important concept.

Very recently, the novel concept of phase transitions has been pioneered \cite{HPK}: the dynamical quantum phase transitions. Below we review the main ideas of this very interesting and quickly developing field of modern quantum physics.

\section{Fisher's zeros} 

Analyzing the behavior of lattice magnetic models in statistical mechanics it is often useful to express the partition function as a polynomial of the temperature or the external magnetic field. The properties of a polynomial are totally determined by the behavior of its roots. This is why, the knowledge of the behavior of zeros of the partition function gives us the possibility to know the total thermodynamics of the studied system. In particular, the knowledge of the distribution of such zeros permits to describe exactly thermodynamics of the considered problem. 

According to the Lee-Yang theorem \cite{LY}  the partition function of a statistical model (with ferromagnetic interactions) is the function of an external magnetic field, then all zeros of the partition function are imaginary (or after the change of variables they are distributed on a unit circle). Consider the Hamiltonian 
\begin{equation}
{\cal H} = -\sum_{j,k} J_{jk}S_jS_k - \sum_j H_j S_j \ , 
\end{equation}
where $S_j$ are Ising spin variables, $J_{jk} >0$ are ferromagnetic coupling constants, and $H_j$ is the external field. The partition function of the system with the Hamiltonian ${\cal H}$ can be written as \cite{New} (here we normalize all values of $J_{jk}$ and $H_j$ by the real positive temperature $T$)
\begin{equation}
Z= \int \exp( -{\cal H} )d\mu_1(S_1) d\mu_2 (S_2)\cdots d\mu_N (S_N) \ , 
\end{equation}
where $d \mu_j$ is the even measure on real $R$, which decreases at infinity, i.e., 
$\int e^{bS^2} d|\mu_j (S) | < \infty$, with any real $b$ belonging to $R$.  Suppose all zeros of the generalized Fourier transformation of the measure $\int e^{HS} d\mu_j(S) \ne 0$ are real for any complex $h$. Then, if the values $z_j$ have the positive real part, the partition function is nonzero, 
$Z (z_j) \ne 0$. In other words, the partition function vanishes for purely imaginary magnetic fields $H_j$. In the Ising model the above mentioned measures are related to the set of values $\pm 1$, so that the partition function can be considered as a function of the variables $x = \exp (\pi z)$. After such a change of variables we see that all zeros of $x$ lie on the unit circle $|z|=1$. If the partition function has no zeros, then the free energy is the analytic function, and corresponding system does not undergo a phase transition. Contrary, if zeros of the partition function do close onto the positive real  axis of $z$, each of such roots would correspond to a discontinuity of the derivative of the free energy, i.e., to the phase transition of the Ising model. The density of roots determines the order of the phase transition. Lee and Yang have generalized this result to any problem of lattice gases with pairwise attraction between particles on a lattice (the condition of the finite even measure at infinity is translated to the infinite repulsion of particles, if two of them occupy the same site). The consequences of the Lee-Yang theorem are, e.g., that the lattice gas cannot undergo more than one phase transition, which corresponds to $z=1$. In particular, for the Ising model the $H$-$T$ phase diagram contains smooth lines, except possibly at zero magnetic field (i.e., at $z=1$). 

In fact, Fisher \cite{Fisher}, instead of the analytic continuation of the external field to complex values, as Lee and Yang did, proposed to study the analytic prolongation of the partition function to complex temperatures $Z (T) \to Z(z)$, where Fisher's zeros are determined from $Z(z_j)=0$. As the size of the system goes to infinity, Fisher's zeros approach the real axis.  

The density of the free energy can be written in the thermodynamic limit $L \to \infty$, where $L$ is the number of sites, as
\begin{equation}
f(z) = -\lim_{L \to \infty} {\ln Z(z)\over L} \ . 
\end{equation}
All contributions to the partition function are terms like $\exp (-zE_j)$, where $E_j$ are eigenvalues of the Hamiltonian, and, hence, they are entire functions of $z$. Therefore, for a finite $L$ the partition function is also an entire function of $z$. According to the Weierstrass factorization theorem \cite{Con} the partition function (as any entire function with zeros) can be written as
\begin{equation}
Z(z) = e^{h(z)} \prod_j \left(1-{z\over z_j}\right) \ , 
\end{equation} 
where $h(z)$ is also the entire function. Therefore, we can write for the density of the free energy
\begin{equation}
f(z) = -\lim_{L \to \infty} {1\over L} \left[ h(z) +\sum_j \ln \left(1-{z\over z_j}\right) \right] \ .
\end{equation}
The non-analytic part of the free energy per site is determined, this way, only by zeroes $z_j$:
\begin{equation}
f^s (z) = -\lim_{L \to \infty} {1\over L} \sum_j \ln \left(1-{z\over z_j}\right) \ . 
\end{equation}
In the thermodynamic limit the sum becomes an integral over some continuous variable $x$ 
\begin{equation}
f^s(z) = -\int_X dx \ln \left(1-{z\over z_j(x)}\right) \ ,
\end{equation}
where $X$ is the region, corresponding to the set of $j$.  After the transformation of the integral we get 
\begin{equation}
f^s(z) = -\int_{z(X)} d{\tilde z} \rho({\tilde z}) \ln \left(1-{z\over {\tilde z}}\right) \ ,
\end{equation}
where $\rho({\tilde z})$ is the Jacobian determinant, which can be considered as the density of zeroes in the complex plane \cite{SK}. It is possible to extend the integration over the full complex plane by setting $\rho(z) = 0$ for $z$ not belonging to $x(X)$. Consider now the real part 
\begin{equation}
\phi(z) = {\rm Re} [f^s(z)] = - \int d{\tilde z} \rho({\tilde z}) \ln |1 - (z/{\tilde z})| \ . 
\end{equation}
For $z= u+iv$ the function $\ln |z|$ is Green's function of the two-dimensional Laplacian $\Delta_{2d} = (\partial^2/ \partial u^2)+  (\partial^2/ \partial v^2)$, i.e., 
 \begin{equation}
 \Delta_{2d} \phi(z) = -2\pi \rho(z) \ . 
\end{equation}  
It means that the density of the free energy can be interpreted as the electrostatic potential $\phi(z)$ produced by the charge density $\rho(z)$ in two dimensions. Hence, the behavior of the free energy at critical points is equivalent to the behavior of the electrostatic potential at surfaces. If zeros form lines in the complex plane, then it is possible to deduce the order of the phase transition directly from the density of zeros \cite{BDL}. Zeros can form areas in the complex plane. These areas can not cover the physical axis for equilibrium phase transitions. However, see below, it is possible for dynamical phase transitions. 

For example, for the Ising model, in the thermodynamic limit Fisher's zeros in the complex temperature plane approach the real axis at the critical value of the temperature $z=1/T_c$. It is the direct indication of the phase transition. For the two-dimensional Ising model Fisher showed that his zeros lie also on a unit circle. However, later it was understood that it is rather exception than the rule, and it is impossible to formulate the analog of the Lee-Yang theorem for Fisher's zeros \cite{SK}. Fisher's zeros form smooth curves, and can, generally speaking, densely occupy entire regions in the complex plane $z$. However, in contrast to the Lee-Yang zeros, the partition function with Fisher's zeros is not a simple polynomial in the complex temperature plane. In a finite system phase transitions cannot occur, and the Fisher zeros are isolated and do not lie on the real axis in the complex temperature plane. However, in the thermodynamic limit Fisher's zeros coalesce into lines or areas, which can cross the real axis. Such a crossing signal  the breakdown of the analytic continuation of the density of the free energy as a function of temperature. 

\section{Loschmidt amplitude} 

The second law of thermodynamics, related to the concept of the time reversal, was the subject of many  discussions. One of them between Joseph Loschmidt and Ludwig Boltzmann, is known as the Loschmidt paradox. Loschmidt pointed out that due to the time-reversal invariance of classical mechanics, an evolution must exist, in which the entropy of the considered system can decrease \cite{Los}. As an example, he considered the case of gas in which all velocities of molecules reverse their sign. Then the entropy of the system would decrease, violating the second law of thermodynamics. Boltzmann answered \cite{Bol} that such a time-reversal is impossible. He pointed out the statistical interpretation of the second law of thermodynamics for generic macroscopic systems. Boltzmann's statement is naturally true for statistical mechanics, however, the Loschmidt paradox can be important for the study of time-reversal dynamics of quantum systems. 

In the isolated quantum system the time evolution is unitary. It implies that if one considers the pure state, then, after the evolution the isolated system remains in the pure state. On the other hand, the connection to the environment yields absence of such a conservation, i.e., to the decoherence. Namely the decoherence, that is related to the interference with the degrees of freedom of the environment of the studied quantum system, produces the classical behavior of the macroscopic systems. The decoherence is the reason for limitation of the use of many quantum systems as elements of a quantum computer \cite{NH}. 

Considering the time evolution of a quantum system it is often instructive to study the behavior of the so-called Loschmidt amplitude (or the Loschmidt echo). It is determined as 
\begin{equation}
G(t) = \langle \Psi_0| e^{i{\cal H}_2t/\hbar}e^{-i{\cal H}_1t/\hbar}|\Psi_0\rangle \ , 
\end{equation}
where $|\Psi_0\rangle$ is the wave function of a quantum system at time $t=0$, ${\cal H}_1$ is the Hamiltonian governing the forward evolution, and ${\cal H}_2$ is the Hamiltonian governing the backward recovery evolution. In such a description the Loschmidt echo defines the degree of (ir)reversibility of the quantum system. From the other viewpoint, it can be considered as the overlap at time $t$ of two time-evoluted wave functions under the effect of two Hamiltonians, ${\cal H}_{1,2}$, i.e. the Loschmidt amplitude  can be considered as the measure for the response of the evolution of a quantum system to perturbations, the fidelity. The Loschmidt amplitude as the measure of the time-reversal, permits to quantify the decoherence effects. The Loschmidt amplitude is usually the decreasing function of time. The main problem is to determine the way of such a decay. It is determined by the most important physical processes, while less important are filtered out by the Loschmidt echo. The return probability (for ${\cal H}_2 = {\cal H}_1$) can be defined as 
\begin{equation}
{\cal L}(t) \equiv |G(t)|^2 = e^{-L l(t)} \ . 
\end{equation}

The Loschmidt amplitude is related to many aspects of research activity of physicists. Here we can mention, e.g., the spin echo in nuclear magnetic resonance \cite{Sli}, consideration of dynamics of nonlinear waves, studies of the quantum chaos, the decoherence in open quantum systems, and statistical mechanics in small systems. It was predicted to be important in many experiments with ultracold atoms trapped in optical cavities \cite{coldthe}. Last but not the least, the Loschmidt amplitude is used in the theory of dynamical quantum phase transitions, which we analyze below. We will mostly consider the response of a quantum system to the global perturbation, i.e., the one, which affects all (or the main part) of the phase space of the system under the time evolution. 

Consider the Loschmidt amplitude for the case in which the backward evolution is excluded, i.e., 
${\cal H}_2=0$ (and ${\cal H}_1 \equiv {\cal H}$). This can be related, e.g., to the situation of the quantum quench, which is the evolution of the quantum system after the sudden change of some parameter(s) of the Hamiltonian of the considered system. The Loschmidt amplitude then describes the evolution of the initially prepared in some initial (pure) eigenstate $|\Psi_0\rangle$ of the Hamiltonian ${\cal H}_0$, under the dynamics, which is regulated by ${\cal H}$ (with the changed parameter). One can see that $G(t)$, i.e., the overlap of the time-evolved quantum state with itself at $t=0$, in this case is similar to the partition function $Z(z) = {\rm Tr} [\exp (-z{\cal H})]$ as the function of the complex temperature, as Fisher suggested, if one considers instead of time the complex variable $z$. We can again determine the density 
\begin{equation}
f(z) = -\lim_{L \to \infty} {\ln G(z)\over L} \  ,  
\label{den}
\end{equation}
where $|\Psi_0\rangle$  play the role of boundary states (for the partition function). We have shown above that the breakdown of the high-temperature expansion of the partition function signals about the temperature-driven phase transition.  Then, the non-analytic time evolution of the Loschmidt amplitude can be considered as the breakdown of the short time expansion at a critical time. It permitted to call such a situation as the dynamical (quantum) phase transition. In most of considered so far cases, the ground state of the quantum system defines $|\Psi_0\rangle$, and, the non-analyticity of the time evolution at the critical time can be determined as the dynamical quantum phase transition. 

Let us return to the analogy between the density of the free energy (now the density of the Loschmidt amplitude) and the behavior of the electrostatic potential in a plane \cite{SK1}. Consider the density of zeros $\rho_1(z)$ in the area I, and the density $\rho_2(z)$ in the area II. At the boundary between areas there is a discontinuous change in the density of  zeros. Let $\phi_{1,2}$ to be the solutions of the Laplace equation with corresponding density. Then consider the behavior of $\phi(z)$ at the intersection of the boundary with the real time axis. Through Stokes' theorem we get
\begin{equation}
{\partial^2 \over \partial y^2} \phi_1(z) = {\partial^2 \over \partial y^2} \phi_2(z) \ , 
\end{equation} 
where  $y$ is the coordinate, parallel to the boundary. Then, we transform to the other set of co-ordinate 
\begin{eqnarray}
&&t= {x\over \cos \alpha} + {y\over \sin \alpha} \ , \nonumber \\
&&y' =y \ , 
\end{eqnarray}
where $\alpha$ is the angle between $t$ and $x$ (the latter is normal to the boundary). Combining it with the Laplace equation we get 
\begin{equation}
\cos^{-2} \alpha {\partial^2 \over \partial t^2} \phi_i(z) + (1+\sin^{-1} \alpha)^2 {\partial^2 \over \partial (y ')^2} \phi_i(z) = -2\pi \rho_i(z) \ , 
\end{equation}
or 
\begin{equation}
{\partial^2 \over \partial t^2} [\phi_1(z)-\phi_2(z)] = -2\pi \cos^2 \alpha [\rho_1(z)-\rho_2(z) ] \ . 
\end{equation}
It means that if the area of zeros of the Loschmidt amplitude overlaps the the real time axis, the second derivative of the real part of the dynamical free energy is discontinuous. 

\section{Dynamical quantum phase transitions in the transverse field  Ising chain}

We start the description of dynamical quantum phase transitions using as an example the one-dimesional chain of Ising spins-1/2 in the transverse magnetic field \cite{HPK}. Such a model often serves as a paradigm for the behavior of quantum phase transitions. The Hamiltonian of the model can be written as 
\begin{equation}
{\cal H} =-2J  \sum_j S^x_jS_{J+1}^x -H\sum_j S_j^z \ , 
\end{equation} 
where $J$ is the exchange constant, $H$ is the external magnetic field, and $S_j^{x,z}$ are operators of projections of a spin 1/2 situated at the site $j$. The Hamiltonian can be diagonalized using the Jordan-Wigner, Fourier, and Bogolyubov transformation. The Jordan-Wigner transformation \cite{JW} yields 
\begin{eqnarray}
&&{\cal H} = -J \sum_j (c_j^{\dagger}c_{j+1} +c_j^{\dagger}c^{\dagger}_{j+1} + {\rm H.c.}) 
\nonumber \\
&&+H\sum_j c^{\dagger}_jc_j -{LH\over 2} \ , 
\end{eqnarray}
where $c^{\dagger}_j$ ($c_j$) creates (destroys) a fermion at the site $j$. Fourier transformation and Bogolyubov transformation yields ${\cal H} =\sum_{k>0} \varepsilon_k [b^{\dagger}_k b_{k} +(1/2)]$, where 
\begin{equation}
\varepsilon_k \equiv \varepsilon_k(H) = \sqrt{(H-J\cos k)^2 +J^2\sin^2 k} \ . 
\end{equation} 
Fermions fulfill either antiperiodic or periodic boundary conditions. Those are usually refereed to as the Neveu-Schwarz, or Ramond sectors, respectively. The momenta $k$ are quantized as either half-integer or integer of $2\pi/L$, respectively, $k=\pi(2n+1)/L$ for antiperiodic, and $k=2\pi n/L$ for the periodic boundary conditions. It is well known that in the ground state the system undergoes the quantum phase transition. For $H < H_c=J$ the Ising chain is ordered with the order parameter $\langle S^x_j\rangle$ \cite{Pfet}. In this phase the transverse field Ising model possesses two degenerate ground states $|\pm \rangle$, in which $\langle S^x_j \rangle \ne 0$. On the other hand, for $H > H_c$ the system is disordered and the ground state is unique. The correlation length diverges at $H_c$ with the correlation function exponent $\nu =1$. Nonzero temperature destroys magnetic ordering, as usual for one-dimensional quantum systems. 
 
Now we are in position to describe the dynamical quantum phase transitions in the transverse field Ising chain. Suppose for $t < 0$ the system was in equilibrium with the Hamiltonian ${\cal H}$ with $H=H_0$. 
Then, at $t=0$ the value of the magnetic field is suddenly changed to $H_1$ (i.e., we have the quantum quench situation). Then the density of the logarithm of the Loschmidt amplitude is the rate function of the return amplitude, $G(t) =\exp [-Lf(z=it)]$. The latter can be rewritten as 
\begin{equation}
f(z) = -{1\over 2 \pi} \int^{\pi}_0 dk \ln \left( \cos^2 \phi_k +\sin^2 \phi_k e^{-2z\varepsilon_k(H_1)/\hbar}\right) \ , 
\label{dens1}
\end{equation} 
where $\phi_k = \theta_k(H_0) -\theta_k(H_1)$, and 
\begin{equation}
\tan [2\theta_k (H)] = {J\sin k\over H-J\cos k}  \ . 
\end{equation}
Here we have dropped the contribution from the ground state (zero oscillations) $(1/2)\sum_k \varepsilon_k(H_1)$, because it obviously does not contribute to dynamics. Notice that the return probability is ${\cal L}(t) =\exp (-L[f(z=it)+f(z=-it)])$. The description above only applies to quenches starting from the unique ground state for the fermionic model, i.e., to the Neveu-Schwarz sector for any finite system. In the paramagnetic phase $H>H_c$ this state corresponds to the superposition of two degenerate ground states $|\pm \rangle$ \cite{Cal}. This is why, the situation with quantum quenches starting from one of the spin-polarized phases $|\pm\rangle$ cannot be described by Eq.~(\ref{dens1}). However, when one starts from either of two degenerate ferromagnetic states, the cusps in the behavior of some time-dependent characteristics of the transverse field Ising chain persist, see below. The Loschmidt amplitude, that way, measures the effect of the quench on the system.  On the other hand, dynamical quantum phase transitions occur whenever the initial state is orthogonal to the time-evolved state after the quantum quench. 

Then we can use the definition of Fisher's zeroes (however for the Loschmidt amplitude). Zeroes of Eq.~\ref{dens1}) coalesce in the thermodynamic limit $L \to \infty$ to the family of lines \cite{HPK}
\begin{equation}
z_n (k) = {\hbar \over 2\varepsilon_k(H_1)} \left[ 2\ln \tan \phi_k +i\pi (2n+1)\right]  \ . 
\end{equation}
It is easy to show that $\phi_{k=\pi} =0$. On the other hand, the value $\phi_{k=0}$ depends on the relative values of $H_0$ and $H_1$. Namely, if both $H_0$ and $H_1$ are within the same phase (i.e., 
both of them are smaller or larger than $H_c$), we have $\phi_{k=0}=0$. If one of fields, say, $H_0 > H_c$, and the other $H_1 < H_c$ (or vice versa) we get $\phi_{k=0} =\pi/2$. This case desribes the quantum quench across the quantum critical point. Finally, if one of $H_0$ or $H_1$ is equal to $H_c$ we get $\phi_{k=0}=\pi/4$. It describes the quench to or from the quantum critical point. 

It means the following. The line of Fisher's zeros cut the time axis (${\rm Re} (z) =0$) for a quench  across the quantum critical point, i.e., $\lim_{k \to 0}{\rm Re} z_n(k)  = \infty$ and $\lim_{k \to \pi}{\rm Re} z_n(k)  = -\infty$. 
\begin{figure}
\begin{center}
\vspace{-15pt}
\includegraphics[scale=0.35]{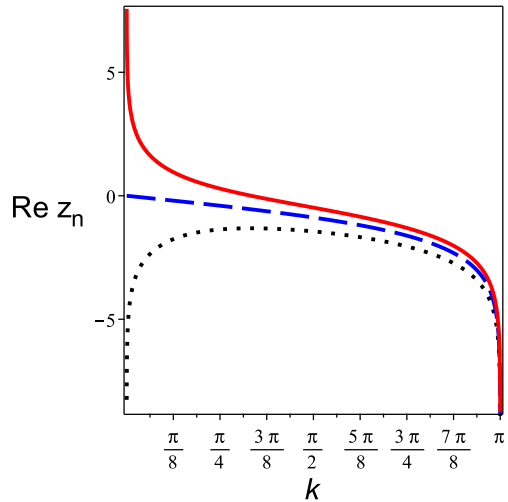}
\end{center}
\vspace{-15pt}
\caption{(Color online) Dependence of ${\rm Re} z_n$ of the transverse Ising chain on the wave vector $k$ for $H_1=0$ and $H_0=0.5H_c$ (black dotted line), $H_0=H_c$ (blue dashed line), and $H_0=2H_c$ (red solid line).}   
\label{Zn}
\end{figure}
This behavior of ${\rm Re} z_n(k)$ for several values of $H_0$ (and $H_1$ fixed) is shown in Fig.~\ref{Zn}. 

It can be shown that such a behavior remains unchanged for any (not sudden) change of the parameter of the Hamiltonian (ramping). Suppose the value $H_1$ is reached at $t=\tau$. Then, for such a general ramping $H(t)$ with $H(t=0)=H_0$ and $H(t=\tau)=H_1$ one can define $|\Psi_0 \rangle = |\Psi (\tau)\rangle$, where $|\Psi (t)\rangle$ is the eigenfunction of the Schr\"odinger equation 
\begin{equation}
i\hbar {\partial |\Psi (t)\rangle \over \partial t} = {\cal H}[H(t)]|\Psi (t)\rangle 
\end{equation}
with the initial condition $|\Psi (t=0)\rangle =|\Psi_0 (H_0)\rangle$. 

\begin{figure}
\begin{center}
\vspace{-15pt}
\includegraphics[scale=0.35]{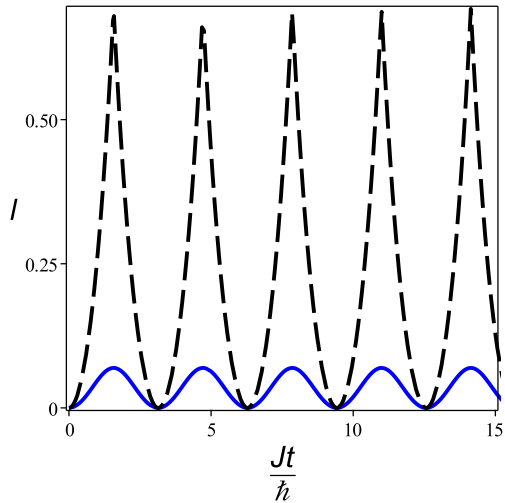}
\end{center}
\vspace{-15pt}
\caption{(Color online) Time dependence of the density of the return probability $l(t)$ of the transverse Ising chain for $H_1=0$ and $H_0=0.5H_c$ (blue solid line) and $H_0=3H_c$ (black dashed line).} 
\label{l}
\end{figure}

For quenches across the quantum phase transition point the non-analytic behavior of the Loschmidt amplitude and the return probability  at special times $t_n^*$ 
\begin{equation}
t_n^* ={\pi \hbar \over \varepsilon_{k^*}(H_1)} \left(n +{1\over 2}\right) = {\pi \hbar \sqrt{H_0+H_1}(2n+1)
\over 2 \sqrt{(H_0-H_1)(J^2-H_1^2)}} \ . 
\end{equation}
follows the behavior of the Fisher zeros \cite{Poll}. Here the value $k^*$ is determined from the condition ${\rm Re} z_n(k^*)=0$, which for the transverse field Ising chain is 
\begin{equation}
\cos k^* = {J^2 +H_0H_1\over J(H_0+H_1)} \ .
\end{equation}
The example of the time dependence of the density of the return probability $l(t)$ is shown in Fig.~\ref{l} for $H_1=0$ and $H_0=0.5H_c$ and $H_0=3H_c$. One can see that the density of the return probability manifests discontinuities (cusps) at special values of $t =t^*$ for the quantum quench across the quantum critical point. It is the manifestation of the dynamical quantum phase transitions. Contrary, for the quantum quench in the same phase, there is a periodicity of the density of the return probability with the period $t^*/\hbar$, however there are no discontinuities and time evolution of the rate function is completely smooth. 
It turns out that the time dependence of Fig.~\ref{l} shows only periodicity, without relaxation. It is the special case, related to the simple limiting case $H_1=0$. Here $\varepsilon_{k^*}(H_1)=J$, and, hence, the real part of the Fisher zeros at $k=k^*$ is equal to zero, ${\rm Re} z_n(k^*) =0$, and, therefore there is no relaxation of $l(t)$. In general, for $H_1 \ne 0$ the time dependence of the density of the return probability reveals periodic oscillations, which decrease with time, due to nonzero ${\rm Re} z_n(k^*)$.   

It is important to stress that the mode $k^*$ is distingushed because in the ground state the occupation of that mode is $n_{k^*}=1/2$ in the basis of eigenstates of the final Hamiltonian ${\cal H} (H_1)$. Modes with $k >k^*$ have the thermal occupation $n_k <1/2$, natural for positive temperatures. However, modes with $k < k^*$ have inverted population $n_k >1/2$, which is reminiscent of the formally negative effective temperature. Hence, the mode with $k^*$ corresponds to the infinite effective temperature.  For any quantum quenches across the quantum critical point such a mode exists, and, hence, the lines of Fisher's zeros cut the time axis (quasi)periodically, implying dynamical quantum phase transitions. The existence of such a mode in relation to spatial correlations was discussed \cite{Kol}. 

In fact, the short time expansion for the rate function $f(z)$ (or for the Loschmidt amplitude $G(z)$ and the return probability ${\cal L}(z)$) is broken down in the thermodynamic limit if there exists a quench across the quantum critical point. It is totally analogous to the breakdown of the high-temperature expansion of the free energy (or the partition function) for the phase transition in equilibrium. It is interesting to notice that for slow ramping $\varepsilon_{k^*}(H_1)$ plays the role of the mass (gap) $m(H_1)=|H_1-H_c|$ of the final state Hamiltonian of the transverse Ising chain. Here the system can be described in terms of massive Majorana fields \cite{Muss}.

That mass is the only energy scale in the equilibrium for the Ising chain for $H=H_1$. On the other hand, this energy is related to the new energy scale, which is generated by the quantum quench. In the vicinity of the quantum critical point of the system, described by the finite Hamiltonian, $H_1 =H_c +\delta$ ($\delta \ll H_c$) one gets that $\varepsilon_{k^*}\propto \sqrt{|\delta|}m(H_1)$, and, therefore, $\varepsilon_{k^*}$ becomes different from the mass (gap). 

\begin{figure}
\begin{center}
\vspace{-15pt}
\includegraphics[scale=0.35]{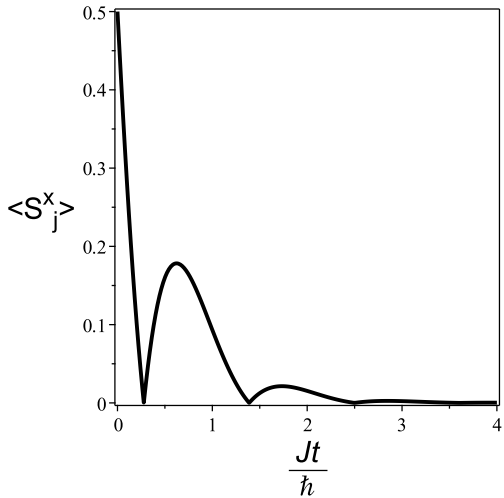}
\end{center}
\vspace{-15pt}
\caption{Time dependence of the density of the order parameter of the transverse Ising chain for $H_0=0$ and $H_1=3H_c$ and $\alpha =\pi/2$.}
 \label{ord}
\end{figure}

Dynamical quantum phase transitions in the transverse field Ising chain manifest themselves also in the time dependence of the order parameter after the quantum quench. For the transverse field Ising chain the ground state order parameter is $\langle S^x_j\rangle$, which is nonzero for $H <H_c$. The calculation of the time dependence of the order parameter was performed in \cite{BM,Cal}. For quenches, starting in the disordered phase $H_0 >H_c$, the order parameter is zero because the symmetry $Z_2$ (between states $|\pm\rangle$) remains unbroken \cite{Cal,Sch}.  According to \cite{Cal,Sch} after the quantum quench, which starts in the ordered phase and finishes in the ordered phase $H_0,H_1 <H_c$, the ground state time dependence of the order parameter is 
\begin{equation}
\langle S_j^x\rangle ={1\over 2} \sqrt{C_1} \exp({t\over \hbar} \int_0^{\pi} dk {H_1\sin k \over \pi \varepsilon_k(H_1)} \ln |\cos |\Delta_k|| \ , 
\end{equation}
where 
\begin{equation}
C_1= {J^2 -H_0H_1 +\sqrt{(J^2-H_1^2)(J^2-H_0^2)}\over 2 \sqrt{J}\sqrt{J^2-H_0H_1}
(J^2-H_0^2)^{1/4}} \ , 
\end{equation}
and
\begin{equation}
\cos (\Delta_k )= {H_0H_1-J(H_0+H_1)\cos k +J^2\over \varepsilon_k(H_0)\varepsilon_k(H_1)} \ ,
\end{equation}
i.e., the order parameter decays with time exponentially. It is clear, because in equilibrium the order parameter exists only for $H<H_c$. On the other hand, for $H_0 < H_c$, and $H_1> H_c$, i.e. for the quantum quench across the quantum critical point, one has \cite{Cal}
\begin{eqnarray}
&&\langle S_j^x\rangle ={1\over 2} \sqrt{C_2} \exp({t\over \hbar} \int_0^{\pi} {H_1\sin k dk\over \pi \varepsilon_k(H_1)} \ln |\cos |\Delta_k|| \nonumber \\
&&\times [1+\cos(2\varepsilon_{k^*}t/\hbar + \alpha) +\dots]^{1/2} \ , 
\end{eqnarray}
where $\alpha$ is some constant and 
\begin{equation}
C_2 = \sqrt{{H\sqrt{J^2-H_0^2}\over J(H+H_0)}} .
\end{equation}
It means that the exponential decay in time is multiplied by the oscillatory behavior with the period of oscillations related to Fisher's zeros. The time dependence of the order parameter after the quantum quench across the quantum critical point is shown in Fig.~\ref{ord}. One can clearly see quasiperiodic cusps related to Fisher's zeros at $t=t_n^*$. 

\section{Full counting statistics approach}

We have shown above the analogy between the equilibrium statistical mechanics and non-equilibrium characteristics of quantum systems. There exists another approach that uses the full counting statistics methods \cite{FCS}. The moment generating function for time-integrated observables is considered as the partition function. Such an approach is known as the $s$-ensemble one, where the counting field $s$ is elevated to that of thermodynamic variable \cite{s}. Similar features in the long time behavior of the cumulant generation function were identified as phase transitions in the full counting statistics \cite{cum}. To remind, the cumulant generating function $K(t)$ is by definition the logarithm of the moment generating function of the random variable $X$, i.e., $K(t) = \ln E[\exp(tX)]$ with cumulants $\kappa_n$ being obtained from the power series expansion $K(t) = \sum_{n=0}^{\infty} \kappa_n t^n/n!$, so that $n$-th cumulant can be obtained by differentiating the above expansion $n$ times and evaluating the result at zero $\kappa_n = K^{(n)}(0)$. Singularities in the cumulant generating function of the Loschmidt amplitude correspond to dynamical quantum phase transitions. We examine the moments of a time-integrated observable of the closed quantum system $Q_t =\int^t dt' q(t')$, where $q(t')$ is the operator, associated with the observable of interest in the Heisenberg representation. Then time evolution operator $T_s (s) =\exp (-it{\cal H}_s/\hbar)$, where $H_s = {\cal H} -isq$ can be used for the moment generating function of $Q_t$ as $Z_t(s) =\langle T_t^{\dagger}(s)T_t(s)\rangle$, and moments are their derivatives, while the logarithm of the moment generating function is the cumulant generating function $A_t(s) =\ln Z_t(s)$. Then we can consider the long-time limit $\alpha(s) = \lim_{L,t\to \infty} [A_t(s)/Lt]$. Let us, following \cite{HGG}  consider the cumulant generating functional for the transverse field Ising model 
\begin{equation}
\alpha(s) = {1\over \pi} {\rm Im} \int_0^{\pi} |[(H+is -J \cos k)^2 +J^2 \sin^2 k]^{1/2}| dk \ . 
\end{equation}
We can consider it as the dynamical free energy of the transverse field Ising model in the complex field $H+is$.  Then it is possible to introduce the dynamical order parameter $-\partial \alpha(s)/\partial s$ and the susceptibility $\chi_s = -\partial^2 \alpha(s)/\partial s^2$. Then the properties of the considered model depend on the parameters $H$ and $s$. We can plot the ground state $H$-$s$ phase diagram where the circle 
\begin{equation}
H^2 +s^2 =J^2 
\end{equation}
determines the quantum critical line, at which the gap of excitations of the the considered model is closed at particular value of $k$. For $|H| < J$ the critical value is given by $s_c=\sin k_H$, with $\cos k_H = H/J$. The region inside the circle is dynamically ordered phase, and the other one is the dynamically disordered phase. One can show that this line corresponds to the second order phase transitions, because the second derivative of $\alpha(s)$ is divergent. At the end points of the quantum critical line at $s=0$ the second order phase transitions appear. Each point of this phase diagram can be associated with the state $|s\rangle = \lim_{t \to \infty} T_t(s) |i\rangle$ for the initial state $|i\rangle$ (for the transverse field Ising model it is the vacuum state) with the necessary normalization. Notice that such a state is independent of the initial state provided the latter has a finite overlap with it.  For $H > J$ the state $|s\rangle$ is proportional to the product over $k>0$ of states with single fermion modes $|1_k\rangle_s$ of the transverse field Ising chain with $s \ne 0$ with $k$ and $-k$. For $H < J$ the state $|s\rangle$ is proportional to the product over $k>0$ of states with zero fermion modes$|0_k\rangle_s$  with $k$ and $-k$. Finally, for $-J < H< J$, it is the product over $k<k_H$ of zero fermion  modes with $\pm k$ times the product over $k>k_H$ of single fermion modes with $\pm k$. The state $|s\rangle$ is the state, which can be obtained under the evolution from $T_t(s)$. The operator $T_t(s)$ can be related to the evolution of the density matrix via the Liouville-like equation as 
\begin{equation}
i\hbar {\dot \rho}(t) =[{\cal H}, \rho(t)] - is\{q,\rho(t)\} \ , 
\end{equation}
which is the Lindblad master equation \cite{Lind}  
\begin{equation}
i\hbar {\dot \rho}(t) =[{\cal H}, \rho(t)]  +i\sum_i \left( L_i\rho L_i^{\dagger} - 
{1\over 2}\{L_i^{\dagger}L_i,\rho\} \right) 
\end{equation}
 with $\sum_i L_i{\dagger} L_i =sq$, and without recycling terms.
For the transverse field Ising chain the Lindblad operator $L_i$ has the meaning of the spin lowering (jump) operator, and $s$ plays the role of damping. The states $|s\rangle$ can be prepared by coupling of the considered system to that simple Markov environment. If the system is evolved without emission, we get the state $|s \rangle$. Then, the quantum quench can be considered as the dynamics of the initially prepared state $|s\rangle$, decoupled from the environment, under the action of the Hamiltonian ${\cal H}$. Similar to the previous section, it is possible to introduce the density of the return probability $l(t)$ for the complex transverse field Ising model. It can be divided into two contributions, for the integrations with respect to $k$: for $0 \le k \le k_H$, and for $k_H \le k \le \pi$. Hence, there are two families of Fisher's zeros. The definition of zeros for the first one is similar to the previous definition. On the other hand, for the second one, it is related to the emergent nonanalytic behavior at the limits $k_H$ of the integrals. In both cases these Fisher's zeros lie on the real time axis   so that the angles $\phi_k^s$ (introduced analogously to the ones in the previous section, but for $s\ne 0$) satisfy the condition $|\cos \phi_k^s| = |\sin \phi_k^s|$. Fisher's zeroes cross the real time axis when the initial $|s\rangle$ lies in the dynamically disordered phase, and $|H| >J$. The occupation mode for $s=0$ of $k_H$ is equal to 1/2. Hence, the mode $k_H$ in this language is analogous to $k^*$.  This mode defines the critical characteristics of the full counting statistics of the time-integrated magnetization of the transverse complex field Ising model. It turns out that dynamical phase transitions in this approach emerge even out of the quantum criticality. 

Now, we can show how the dynamical properties of the system out of equilibrium can be connected to the topology. Topological quantum numbers provide the way for characterization of the ground state properties of many-body quantum systems \cite{KGP}, and new geometric interpretation of quantum phase transitions \cite{tqp}. 

One of the generally used measures of geometric properties of the system is the Pancharatnam-Berry phase \cite{Berr} (let us call it just Berry phase below, following the used practice).  It is the phase difference acquired over the course of a cycle, when a considered system is subjected to cyclic adiabatic processes. It results from the geometrical properties of the parameter space of the Hamiltonian of the system. For example, let us consider the manifold of Hamiltonians defined with some parameters ${\bf \lambda}$. The natural measure of the distance between the ground states $|0({\bf \lambda})\rangle$ of this manifold \cite{PV} is 
\begin{equation}
1 - |\langle 0({\bf \lambda})|0({\bf \lambda} +d{\bf \lambda}) \rangle|^2 = \sum_{\mu,\nu} g_{\mu \nu} d\lambda^{\mu} d\lambda^{\nu} \ , 
\end{equation}
where $g_{\mu \nu}$ is the geometrical tensor 
\begin{eqnarray}
&&g_{\mu \nu} = \langle 0({\bf \lambda})|\partial_{\mu} \partial_{\nu} |0({\bf \lambda} \rangle - 
\nonumber \\
&&\langle 0({\bf \lambda})|\partial_{\mu} |0({\bf \lambda} \rangle \langle 0({\bf \lambda})|\partial_{\nu} |0({\bf \lambda} \rangle \ , 
\end{eqnarray}
where $\partial \mu \equiv \partial/\partial \lambda^{\mu}$, and the partial derivatives for $\mu$ and $\nu$ act to the left, and to the right, respectively. The Berry curvature $F_{\mu \nu}$ is related to the imaginary part of the geometric tensor
\begin{equation}
F_{\mu \nu} = -2 {\rm Im} [g_{\mu \nu} ] = \partial_{\mu} A_{\nu} - \partial_{\nu} A_{\mu} \ , 
\end{equation}
where $A_{\mu} = i\langle 0({\bf \lambda})|\partial_\mu|0({\bf \lambda})\rangle $ is the Berry connection. The Berry phase ${\cal B}$ is the line integral 
\begin{equation} 
{\cal B} = \int_{\partial S} {\bf A} d{\bf \lambda} \ . 
\end{equation} 
The transport along a two-dimensional manifold leads to the Chern number of the system; it is related to the surface integral of the Berry curvature 
\begin{equation}
{\cal C} = {1\over 2 \pi} \int_M F_{\mu \nu} dS_{\mu \nu} \ . 
\end{equation}
If the manifold is closed then the Chern number is integer \cite{Tho}. 

Now, consider the evolution of the ground state under the quantum quench from the initial state $\lambda_i=H_0$ with the Hamiltonian determined by $\lambda_f=H_1$. The time evolution for the transverse field Ising model factorizes into contributions from each $k$-sector, and the wave function is the Cooper pair-like one 
\begin{equation} 
|u_{k,t}\rangle = [\cos \phi_k -i e^{-2i\varepsilon_k(H)t/\hbar} \sin \phi_k b^{\dagger}_k b^{\dagger}_{-k}] |0\rangle \ . 
\end{equation}
Notice, that the time evolution of the states does not obey the global $U(1)$ symmetry (up to the multiplier $\exp (-i 2\varphi)$), while the Hamiltonian does. For the manifold of values of $k$ and $\varphi$ the Berry curvature and the Berry phase are time-independent, although quenched states depend on time. The states are uniquely defined for $0 \le \varphi , k \le \pi$. The excited states are orthogonal to the ground state. The states at $k=\pi$ do not depend on $\varphi$, independent on the quench ramping protocol. For $k \to 0$ the behavior is more complex. Namely, within the same phase $k=0$ states are $\varphi$-independent,  $\sin \phi_{k \to 0} =0$, and, hence, the considered manifold is equivalent to the $S^2$-sphere. However, if we consider the quench across the quantum critical point, we have $\sin \phi_{k \to 0} =1$ and states depend on $\varphi$ (up to a global phase, which can be removed by the gauge transformation), which again leads to the $S^2$-sphere. This is why, the critical dynamical phase transitions correlate with the need to gauge fix the $k=0$ modes when considering the topology of the manifold. This, in turn, alters the Chern number associated with the manifold of quenched states. One can choose ${\cal C} = \sin^2 \phi_k$. For quenches within the same phase we have ${\cal C}=0$, and for quantum quenches across the quantum critical point we have ${\cal C}=1$. 

In the full counting statistics approach one can use the states $|s_k\rangle$ ($|s\rangle = \otimes_{k>0} |s_k \rangle$, and $|s_t\rangle = \exp (-i{\cal H} t/\hbar)|s \rangle$) instead of the states  $u_{k,t}\rangle$.  The geometry of the ground state of the system manifests the signatures of the quantum critical line. Diagonalizing both ${\cal H}_s$ and ${\cal H}$ one can express the state $s_t\rangle$ in terms of fermionic modes of the final Hamiltonian ${\cal H}$ of the transverse field Ising chain. Then, after the gauge transformation, we get $|s(\varphi)\rangle$, with the Berry phase  
${\cal B} =i \int_0^{\pi} d\varphi \langle s(\varphi)|\partial_{\varphi}|s(\varphi)\rangle$. In the thermodynamic limit we get \cite{HGG}
\begin{eqnarray}
&&b=- \lim_{L \to \infty} {{\cal B}\over L} = - \int_{k_H}^{\pi} dk {|\cos \phi_k^s|^2 \over \cosh (2{\rm} \alpha_k^s)}  - \nonumber \\
&&\int_{0}^{k_H} dk {|\sin \phi_k^s|^2 \over \cosh (2{\rm} \alpha_k^s)}  \ . 
\end{eqnarray}
We can check that $b$ is the same for the dynamically ordered and disordered phases. However, $db/ds$ shows minima at the full counting statistics critical line. Also, the Chern number can be obtained \cite{HGG}. It has a nonanalytic point exactly at the quantum critical line of the full counting statistics. In summary, dynamical quantum phase transitions in the full counting statistics approach can exist even if in the equilibrium there are no quenches across quantum  phase transitions. 

It is interesting to notice that the recent study \cite{Hic} , which uses similar approach, came to the conclusion that dynamical quantum phase transitions in the XY spin-1/2 model are related to the third order of the equilibrium thermodynamics.  Also interesting, that  dynamical quantum phase transitions in the Hubbard model \cite{Hub} and in the Falicov-Kimball model  \cite{FK} were shown to be the first order transitions (i.e., co-existing of two solutions)  \cite{CWE}, using the non-equilibrium dynamical mean-field theory \cite{DMFT} based on the many-body Keldysh formalism \cite{Kel}, i.e., in the general framework for describing the quantum mechanical evolution of a system in a non-equilibrium state.  It can be compared with {\em jumps} of the order parameter at critical times for the transverse field Ising chain \cite{HPK}, see also \cite{KS,KKK} for the non-integrable models. Note, however, that \cite{VD2} mentioned  the different order of the dynamical quantum phase transitions.     

\section{Generalization to other one-dimensional integrable models}

It is possible to generalize the above presented results and concepts to other one dimensional exactly solvable models. 

\begin{figure}
\begin{center}
\vspace{-15pt}
\includegraphics[scale=0.35]{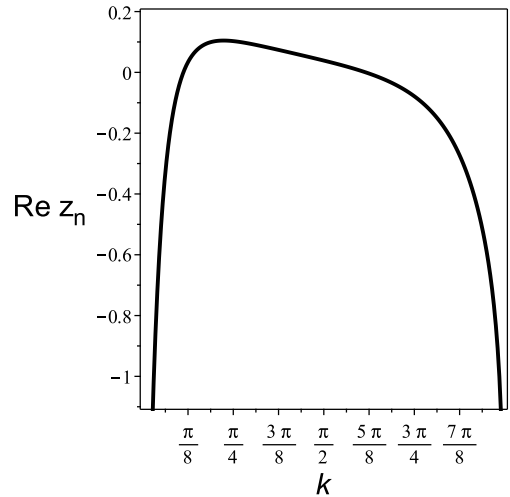}
\end{center}
\vspace{-15pt}
\caption{Dependence of ${\rm Re} z_n$ of the XY spin-1/2 chain on the wave vector $k$ for $H_1=1.5H_c$ and $H_0=2H_c$, $\gamma_0=-5$ and $\gamma_1 =1$.}   
\label{Zn1}
\end{figure}

The simplest generalization of the transverse field Ising chain is the XY spin-1/2 one-dimensional model \cite{LSM}. The Hamiltonian of the model is  
\begin{equation}
{\cal H} = \sum_j (2J_xS_j^xS_{j+1}^x +2J_yS_j^yS^y_{j+1} -HS_j^z) \ , 
\end{equation}
The Hamiltonian is diagonalized using the Jordan-Wigner, Fourier, and Bogolyubov transformations, similar to the transverse field Ising chain. The difference is in the dispersion relation 
\begin{equation}
\varepsilon_k = \sqrt{(H-J \cos k)^2 +J^2\gamma^2 \sin^2 k} \ , 
\end{equation}
where $J=J_x+J_y$, and  $\gamma = (J_x-J_y)/J$ defines the anisotropy of the exchange couplings in the XY plane. The dispersion relation is gapped both for $H < H_c$ and $H >H_c$, and it is equal to zero for $H=H_c$ only at $k=0$. It is obviously gapless for $\gamma =0$ for $H <H_c$. In this case the equilibrium quantum phase transition takes place from the disordered gapless phase at $H < H_c$ to the spin-polarized phase at $H > H_c$, where $S^z_j=1/2$. The gap is equal to $H-|J|$ for $H >  |J||(1-\gamma^2)|$, and $|\gamma |\sqrt{J^2(1-\gamma^2) -H^2}/\sqrt{1-\gamma^2}$ otherwise. The Hamiltonian again conserves the parity of $S^z +L/2$, where $S^z$ is the $z$-projector of the total spin of the system. The ground state is unique in a given subspace, however for $|H| <H_c=J$ the ground states with even and odd parities are degenerate in the thermodynamic limit. In the even and odd subspaces of $S^z +L/2$ the Loschmidt amplitude gets additional factor $\exp (i t [\pm \varepsilon_{k=0} \pm \varepsilon_{k=\pi}]/2\hbar)$ in the odd sector. Here signs are determined from the fact, in which phase the system was before and after quenches. The main difference of the XY model comparing to the transverse field Ising model is in the second governing parameter, $\gamma$. It implies different definitions of the parameters $\tan 2\theta_k (H) = J\gamma \sin k/[H-J\cos k]$, which renormalizes the value of the real part of $z_n(k)$ and $k^*$. It was pointed out that in the XY model there exists the situation, in which dynamical quantum phase transitions can manifest themselves without crossing critical points of the equilibrium \cite{VD1}. It is possible to show \cite{VD1} that dynamical phase transitions exist if both sets of governing parameters $(H_0,\gamma_0)$ and $(H_1,\gamma_1)$ are inside the phase with $H >H_c$, and $2\gamma_0 \gamma_1 J^2 < J^2 -H_0H_1 -\sqrt{(H_0^2-J^2)(H_1^2-J^2)}$. We can illustrate it in Fig.~\ref{Zn1}, where the real part of $z_n(k)$ is plotted. We can see that the curve $z_n(k)$ crosses the line of imaginary time, which signals about the dynamical quantum phase transition. 
The curve ${\rm Re} z_n (k)$ crosses the imaginary axis twice, which implies two non-equilibrium time scales in the behavior of the density of the Loschmidt amplitude (the dynamical free energy) due to Fisher's zeros. 

\begin{figure}
\begin{center}
\vspace{-15pt}
\includegraphics[scale=0.35]{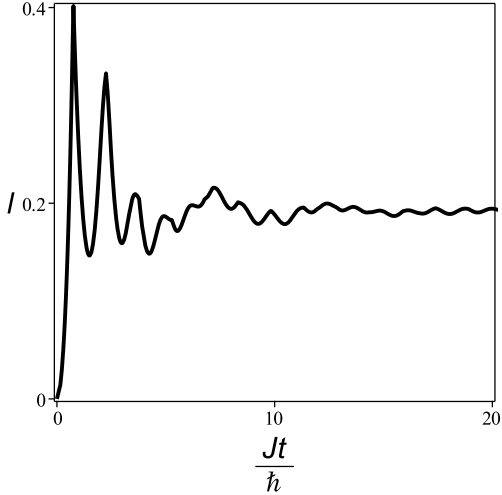}
\end{center}
\vspace{-15pt}
\caption{Time dependence of the density of the return probability $l(t)$ of the XY spin-1/2 chain for $H_1=1.5H_c$ and $H_0=2H_c$, $\gamma_0=-5$ and $\gamma_1 =1$.}   
\label{l1}
\end{figure}

Obviously for $\gamma_0 =0$ there are no dynamical phase transitions. However, the equilibrium quantum phase transition exists at $H=H_c$. On the other hand,  for the quench from the phase with $\gamma_0 \ne 0$ to the phase with $\gamma_1=0$ the dynamical phase transitions can exist.  
For the quenches with the change of the sign of the anisotropy parameter $\gamma$, again, the  curves ${\rm Re} z_n (k)$ cross the imaginary axis twice with two emergent time scales for the behavior of the density of the Loschmidt amplitude. The time dependence of the density of the return probability $l(t)$ is presented in Fig.~\ref{l1}. One can see that while the quench exists in the same equilibrium phase, the density clearly reveals dynamical quantum phase transitions. 
The time dependence of the ground state order parameter for the XY model (magnetization in the XY plane) also follows the above mentioned rules. 

The Hamiltonian of the spin-1/2 XY chain after the Jordan-Wigner transformation is equivalent to the Hamiltonian of the Kitaev one-dimensional model of the topological superconductor \cite{Kit}. Hence, dynamical quantum phase transitions are characteristic for one-dimensional topological superconductors, too \cite{Zv1}. 

It is important to point out that it is possible to have two series of Fisher's zeros. Consider, for example, the Kitaev chain with the hopping between nearest neighbors and additional hopping between next-nearest neighbors \cite{BH}. Then, if we determine the ratio of those two hopping amplitudes as $r$, for the quantum quench from the state with the hopping and pairing amplitudes equal to zero to  the state with nearest and next nearest hopping, two critical momenta $k^*_1 =\pi/2$ and $k_2^* = \arccos (-1/2r)$ can exist. It follows that the time dependence of the density of the return probability shows double (quasi)-periodicity due to such two series of Fisher's zeros \cite{BH}. 

Similar dynamical quantum phase transitions after quantum quenches were studied theoretically in the spin chain in the staggered magnetic field \cite{AS,Zv}, in the dimerized spin chain \cite{Zv}, in the one-dimensional topological insulator in the external magnetic field, in the semiconductor wire with the spin-orbit interaction in the tilted magnetic field  \cite{Zv}, and in the dimerized spin chain with three-spin coupling in the staggered magnetic field \cite{DSD}.

\begin{figure}
\begin{center}
\vspace{-15pt}
\includegraphics[scale=0.35]{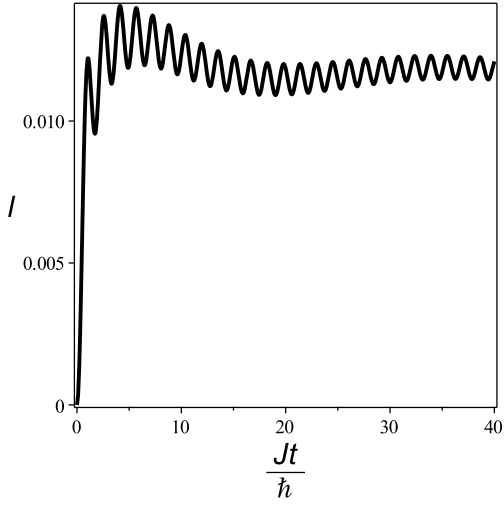}
\end{center}
\vspace{-15pt}
\caption{Time dependence of the density of the return probability $l(t)$ of the XX spin-1/2 chain for 
$\mu_1=-\mu_2$, and $2\mu_1H_0 =0.1J$ and $H_1=0$ .}   
\label{lstag}
\end{figure}

\begin{figure}
\begin{center}
\vspace{-15pt}
\includegraphics[scale=0.35]{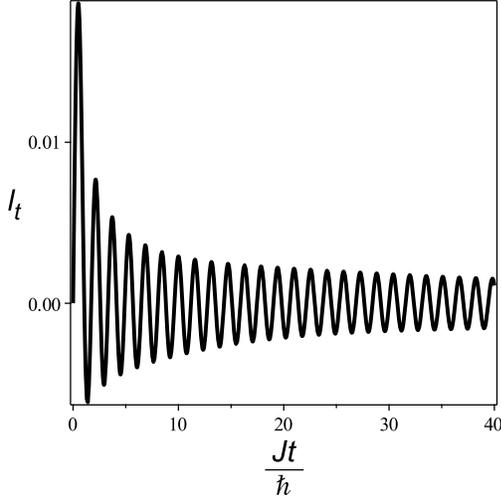}
\end{center}
\vspace{-15pt}
\caption{Time dependence of the time derivative $l_t$ of the density of the return probability $l_t$ of the XX spin-1/2 chain. The parameters are the same as in Fig.~\ref{lstag}.}   
\label{ltstag}
\end{figure}

\begin{figure}
\begin{center}
\vspace{-15pt}
\includegraphics[scale=0.35]{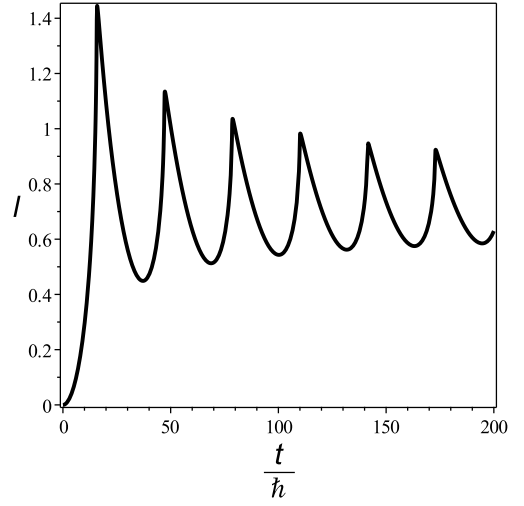}
\end{center}
\vspace{-15pt}
\caption{Time dependence of the density of the return probability $l(t)$ of the XX spin-1/2 chain for 
$\mu_1=-\mu_2$, and $2\mu_1H_0 =200J$ and $H_1=0$ .}   
\label{las}
\end{figure}

\begin{figure}
\begin{center}
\vspace{-15pt}
\includegraphics[scale=0.35]{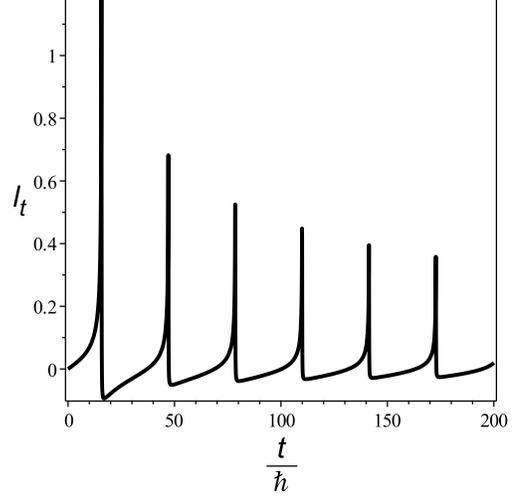}
\end{center}
\vspace{-15pt}
\caption{Time dependence of the time derivative $l_t$ of the density of the return probability $l_t$ of the XX spin-1/2 chain. The parameters are the same as in Fig.~\ref{las}.}   
\label{ltas}
\end{figure}

Let us consider, for example the dimerized XX spin-1/2 chain in the homogeneous and staggered magnetic field. The Hamiltonian of the latter has the form \cite{Zvbook} 
\begin{eqnarray}
&&{\cal H}_{d} = \sum_{n}\biggl[ J_1 (S_{1,n}^xS_{2,n}^x + S_{1,n}^yS_{2,n}^y)+ J_2(S_{1,n+1}^x \nonumber \\
&&\times S_{2,n}^x +S_{1,n+1}^yS_{2,n}^y) - H(\mu_1 S_{n,1}^z + \mu_2S_{n,2}^z) \biggr] \ , 
\end{eqnarray}
where $S_{n,1,2}^{x,y,z}$ are operators of the projections of the spin at the site $n$ (all spins are divided into two sublattices 1 and 2, and spins, belonging to the first sublattice interact with nearest neighbors from the second sublattice), $J_{1,2}$ describe the exchange integrals, $H$ is the magnetic field, and $\mu_{1,2}$ are the effective magnetons of spins for each sublattice. After the Jordan-Wigner \cite{JW} Fourier, and Bogolyubov transformations the Hamiltonian ${\cal H}_{d}$ can be written as ${\cal H}_d =\sum_k \sum_{1,2} \varepsilon_{1,2,k} c_{1,2,k}^{\dagger}c_{1,2,k} +{\rm const.}$, where
\begin{eqnarray}
&&\varepsilon_{1,2,k} = H{(\mu_1+\mu_2)\over 2} \pm 
\nonumber \\
&&{1\over 2}[H(\mu_1-\mu_2)^2 + 4J_1^2+4J_2^2 +8J_1J_2\cos k]^{1/2}  \ . 
\end{eqnarray}
The fermionic Hamiltonian of the dimerized spin chain is equivalent to the SSH model, introduced to model poliacetilene \cite{SSH}.  There are two critical fields $H_{1,2}=|J_1\pm J_2|/2\sqrt{|\mu_1\mu_2|}$ at which quantum phase transitions take place. In the ground state for $H \ge H_2$ the spin chain is in the spin-saturated phase with the nominal total magnetization per lattice unit $(\mu_1+\mu_2)/2$. For $H \le H_1$ the $z$-projection of the total spin moment is zero. Such features of the behavior of the static characteristics determine the ground state behavior of the Loschmidt amplitude. In the ground state the integration with respect to $k$ for $H \le H_1$ is taken over all $k$, $\sum_k  \to (L/\pi)\int_0^{k_0} dk$ with $k_0=\pi$. On the other hand, for $H \ge H_2$ we have $k_0=0$ and the change of the magnetization of the dimerized spin chain in the ground state is zero. Finally, for $H_1 < H < H_2$ the integration limit is $k_0 = \arccos [(\mu_1\mu_2H^2 -J_1^2 -J_2^2)/2J_1J_2] = \arccos [(2H^2-H_1^2-H_2^2)/(H_2^2-H_1^2)]$. Let us consider the quantum quench for the case $\mu_1 =-\mu_2$, i.e., only staggered component of the magnetic field is present, and $J_1=J_2=J$ (no dimerization) (cf. \cite{AS}). Let the initial state be the state with $H_0 \ne 0$, and the final state is $H_1=0$. The results for the density of the return probability $l(t)$ and its time derivative $l_t(t) \equiv dl(t)/dt$ are presented in Figs.~\ref{lstag} and \ref{ltstag}, respectively. 
We do not see dynamical quantum phase transitions in this model for such a choice of parameters, despite $H$  starts from the equilibrium quantum critical point $H=0$.  It is important to notice that the state with $H_0 \gg J$ is related to the N\'eel state \cite{AS}. In Figs.~\ref{las} and \ref{ltas} the time dependence of the density of the return probability and its time derivative is shown for the case of the quantum quench from the state $J=0.1$ and $H_0=20$ to the state with $H_1=0$. 
In this limit the features in the time dependence of the Loschmidt echo, i.e., the manifestations of the dynamical quantum phase transitions, are clearly seen. The period of oscillations can be approximated as $t_0(k^*) \approx  \hbar \pi / 2J$. 

Now, let us estimate, how the interaction between fermions can affect the dynamical quantum phase transitions after quantum quenches. As an example, let us consider the XXZ spin-1/2 chain with the Hamiltonian 
\begin{equation}
{\cal H} = \sum_j J(S_j^xS_{j+1}^x +S_j^yS_{j+1}^y +\Delta S_j^zS_{j+1}^z)  -H S_j^z \ . 
\label{XXZ} 
\end{equation} 
It is known that depending on the values of $\Delta$ and $H$ the ground state of the model is in the ferromagnetic state (with gapped excitations), in the antiferromagnetic state (with gapped excitations), and in the Luttinger liquid state with gapless excitations. For $H=0$ the quantum critical points are $\Delta =\pm 1$ (for $\Delta=1$ the quantum phase transition is of the Berezinskii-Kosterlitz-Thouless type \cite{BKT}). It turns out that the chain is integrable, i.e. , it has the infinite number of conservation laws \cite{Zvbook1}. These conservation laws affect the thermalization of the quantum system at long times after the quantum quench. Let us consider the Luttinger liquid phase in the bosonization technique \cite{Zvbook}; in the low-energy sector the Hamiltonian can be written as  \cite{AS}
\begin{eqnarray}
&&{\cal H} \approx v_F \sum_{k >0} k \biggl[ \left( (1+{g_4\over 2\pi v_F} \right) (b^{\dagger}_{R,k}b_{R,k} +b^{\dagger}_{L,k} b_{L,k}) + 
\nonumber \\
&&{g_2\over 2\pi v_F}(b^{\dagger}_{R,k}b^{\dagger}_{L,k} +b_{L,k} b_{R,k}) \biggr] \ , 
\end{eqnarray}
where $b^{\dagger}_{R,L,k}$ ($b_{R,L,k}$) create (destroy) the right or left moving bosonic collective excitations (particle-hole pairs), and $g_{2,4}$ are forward scattering amplitudes. The Hamiltonian can be diagonalized by the Bogolyubov transformation 
\begin{eqnarray}
&&{\cal H} = (1/2\pi) \sqrt{[2\pi v_F-g_4]^2 -g_2^2} \nonumber \\
&&\times \sum_k k(\alpha_{R,k}^{\dagger}\alpha_{R,k} +\alpha_{L,k}^{\dagger}\alpha_{L,k}) \ .  
\end{eqnarray}
It is possible to introduce the Luttinger liquid exponent $K$ via ($v= \sqrt{[2\pi v_F-g_4]^2 -g_2^2}/2\pi$) 
\begin{equation}
{2\pi v_F +g_4\over 2\pi v} = {2(1+K)^2\over 4K} -1 \ . 
\end{equation}
The velocity $v_F$ and the Luttinger liquid parameter $K$ can be extracted from the exact solution for the XXZ spin-1/2 chain for $H=0$ as 
\begin{eqnarray}
v_F = {\pi \sqrt{1-\Delta^2}\over 2\arccos \Delta}  \ , \nonumber \\
K= {\pi \over 2(\pi - \arccos \Delta)} \ . 
\end{eqnarray}
It is possible to approximate the values of $v_F$ and $K$ for $\Delta =1$ and $H \le H_c =2J$ \cite{Zv10}. 
The density of the return probability can be written as \cite{AS,ret}
\begin{equation}
l(t) = - {1\over 2\pi} \int_{0}^{\Lambda} dk \ln \left( {1-{\bar K}^2\over 1+{\bar K}^4 -2{\bar K}^2 \cos  (2tvk)}\right) \ , 
\end{equation}
where $\bar K = (K_0-K_1)/(K_0+K_1)$, where $K_{0,1}$ are the Luttinger liquid parameters before and after the quantum quench, and $\Lambda$ is the cut-off. Obviously, there are no Fisher's zeros. 
We have to point out that the time dependence (including the problem of quantum quenches) is not the low-energy problem, as the Luttinger liquid approximation is. The Luttinger liquid approximation is, therefore,  only valid for small quenches, where a small amount of energy is put into the system.  

Comparison of the results of the analytic Luttinger liquid approach and numerical light cone renormalization group algorithm \cite{ES} yields the following \cite{AS}. Both approaches show that there are no dynamical quantum phase transitions for quenches inside the Luttinger liquid phase, while there exist oscillations of the density of the return amplitude with time. However, the quantum quenches across the transitions between the Luttinger liquid and the gapped ferromagnetic and antiferromagnetic phases (including the case of the Berezinskii-Kosterlitz-Thouless point) do show dynamical quantum phase transitions, according to the numerical calculations \cite{AS}. The quenches inside the antiferromagnetic phase, again, do not show dynamical phase transitions. This is why, we can conclude that interactions in integrable one-dimensional spin models do not destroy dynamical quantum phase transitions. 

Notice, that the Loschmidt echo can be described for integrable systems using the thermal Bethe ansatz technique \cite{Zvbook1}, which uses essentially the Trotter-Suzuki decomposition of the quantum transfer matrix \cite{TS}. In that approach the density of free energy can be considered in the thermodynamic limit as the logarithm of the lowest eigenvalue $\Lambda_0$ of the quantum transfer matrix 
\begin{equation}
f(z) = \ln (\Lambda_0(z)) \ . 
\end{equation}
$\Lambda_0(z)$ can be calculated in the framework of the thermal Bethe ansatz (quantum transfer matrix approach) \cite{Zvbook1}. It is possible to introduce the analogy between the complex temperature and the complex time, as before, and, therefore, to establish the connection  between the density of the free energy and the density of the Loschmidt amplitude. Instead of the torus boundary conditions in equilibrium thermodynamics, one has to consider a cylinder with boundaries fixed by the initial state $|\Psi_0\rangle$. There can be no gap between the lowest and the next-to lowest eigenstate $\Lambda_1(z)$ of the quantum transfer matrix. Such level crossing makes $f(z)$ a nonanalytic function. The values of $z=z_n$ at which $\Lambda_0(z_n) =\Lambda_{j\ne 0}(z_n)$ can be considered as Fisher's zeros. Notice that only true crossing (not a degeneracy) leads to Fisher's zeros. On the other hand, if we remember the definition of the correlation length $\xi$ in the quantum transfer matrix approach, 
\begin{equation}
\xi^{-1} = \ln( |\Lambda_0/\Lambda_1|) \ , 
\end{equation}
it is clear that the crossing implies a divergent correlation length $\xi(z_n)$ at Fisher's zeroes, indicating, this way, the dynamical quantum phase transition. It is also possible to define other correlation lengths $\xi^{-1}_j =\ln (|\Lambda_0/\Lambda_j|)$ with similar properties. 

It is worth to mention the important thing. For a finite system the Loschmidt amplitude in the spectral representation can be written as 
\begin{equation} 
G(z) = \sum_n |\langle \Psi_0 |n\rangle|^2 e^{-zE_n}  = {\rm Tr} (\rho_{diag} e^{-z{\cal H}}) \ , 
\end{equation}
where ${\cal H}|n\rangle =E_n|n\rangle$, and the density matrix of the diagonal ensemble is defined as 
\begin{equation}
\rho_{diag} =\sum_n  |\langle \Psi_0 |n\rangle|^2 |n\rangle \langle n| \ .
\end{equation}
Here the Loschmidt amplitude is not a simple fidelity between the initial and time-evolved state, but rather involves a sum over all eigenstates of the final Hamiltonian. 
 It is possible to replace the diagonal ensemble by the generalized Gibbs ensemble, which is the function of the local conservation laws \cite{RDYO}. For the non-integrable system the latter reduces to the canonical or grand canonical ensemble. The numerial calculation \cite{AS} uses the diagonal ensemble. On the other hand, in \cite{Fag} the non-analytic behavior of the Loschmidt amplitude, i.e., dynamical quantum phase transitions, were studied using the exact quantum transfer matrix (thermal Bethe ansatz) approach with the help of the generalized Gibbs ensemble for quenches of $\Delta$ in the XXZ spin-1/2 chain. However, dynamical quantum phase transitions were shown to exist not for all quantum quenches. 

\section{Two-dimensional integrable models}

Dynamical quantum phase transitions were studied also in two-dimensional systems. For example, 
\cite{JK}  studied quantum quenches in the two-dimensional Ising model using the matrix product state approach \cite{MPS} via the numerical algorithm similar to the density matrix renormalization group approach \cite{KA}. The two-dimensional Ising model was presented as a coupled set of one-dimensional Ising chains in the continuum limit. It was shown that the oscillations of the Loschmidt amplitude revealed monotonic oscillations  in the time-dependence for $J_{\perp} \le  0.27 J$, where $J$ is the exchange coupling along chains, and $J_{\perp}$ is the coupling constant between chains. On the other hand, for  $J_{\perp} >  0.27 J$ non-analyticities, i.e., dynamical quantum phase transitions were seen. Notice that the value, at which dynamical quantum phase transitions can exist is larger than the critical value $J_{\perp,c} =0.185 J$, which exists in the equilibrium \cite{KA}.  

The other two-dimensional model, in which dynamical quantum phase transitions were studied, is the honeycomb Kitaev spin-1/2 model \cite{Kithon} with the Hamiltonian 
\begin{equation}
{\cal H}_{Kit} =-4\sum_{\alpha=x,y.z} J_{\alpha} \sum_{\alpha-{\rm links}}S_j^{\alpha}S_{j'}^{\alpha} \ , 
\label{Hxyz}
\end{equation}
where $J_{x,y,z}$ are exchange integrals (let us for definiteness consider the case with $J_x,J_y,J_z \ge 0$), and $S^{x,y,z}_j$ are operators of spin projectors of the spins situated at the sites $j$ of the lattice. Spins interact if they are situated at the neighboring sites. The special feature of the Kitaev model is that the interactions depend on the link type (i.e., along the links parallel to $z$ axis only $z$-projections of spins interact, etc.) \cite{Kithon}. It is possible to use the integrability of the Kitaev honeycomb model. We rewrite the Hamiltonian ${\cal H}_{Kit}$ exactly using the transformation to fermion operators of creation and destruction $d_{j}^{\dagger}$ and $d_j$ (for our purpose it is convenient to use the Dirac representation for fermion operators; however in general \cite{Kithon} the Majorana representation is also useful)). It is the two-dimensional generalization \cite{a,SK1} of the Jordan-Wigner transformation \cite{JW}. The Hamiltonian ${\cal H}_{Kit}$ can be written as the Hamiltonian of the Fermi gas on a square lattice (the Kitaev honeycomb model is equivalent to the brick model \cite{a}) with the site-dependent chemical potential 
\begin{eqnarray} 
&&{\cal H}_{Kit} = 4 \sum_j \biggl[ J_x (d_j^{\dagger} +d_j)(d_{j+{\hat x}}^{\dagger} - d_{j+{\hat x}})  +J_y  
\nonumber \\
&&\times (d_j^{\dagger} +d_j)(d_{j+{\hat y}}^{\dagger} - d_{j+{\hat y}}) + 
2J_z \alpha_j (2d^{\dagger}_j d_j  -1)\biggr] \ , 
\end{eqnarray}
where $\alpha_j =\pm 1$ ($\alpha_j$ commutes with $d_{j'}$ and $d_{j'}^{\dagger}$ for any $j$ and $j'$). This transformation is exact (unlike the approximate Holstein-Primakoff one \cite{HP}), and it is valid for any $J_x,J_y,J_z$. In the sectors with fixed $\alpha_j$ the diagonal form of the Kitaev model can be obtained after the Fourier and Bogolyubov transformations. It has the Bardeen-Cooper-Schrieffer form \cite{BCS} with the energy $E_{\bf k} =4\sqrt{\epsilon_{\bf k}^2 +\Delta_{\bf k}^2}$, where $\epsilon_{\bf k} =\pm J_z +J_x \cos k_x +J_y \cos k_y$ and $\Delta_{\bf k} = J_x \sin k_x +J_y \sin k_y$. The spectrum is gapless for $|J_x-J_y| \le J_z \le J_x +J_y$, and gapped otherwise. The summation is over all ${\bf k}$ belonging to the subset of the Brillouin zone such that $- {\bf k}$ is out of that subset \cite{SK1}.

Dynamical quantum phase transitions were studied in the Kitaev honeycomb spin model in \cite{SK} for the sector with with all $\alpha_j=-1$ \cite{Kithon,Lieb,SK1}. The study was similar to the approach of \cite{HPK} for the transverse Ising chain, because in the latter the Hamiltonian also had BCS-like form. The principal difference is in the dimensionalilty. The quantum quench was from the initial state with the one set of exchange constants ${\bf J}_0$ to the final state with the other set ${\bf J}_1$, where ${\bf J}_{0,1} = (J_x,J_y,J_z)_{0,1}$ before and after the quantum quench, respectively. The partition function $Z(z)$ can be expressed \cite{SK1} via the BCS eigenfunction 
\begin{equation}
Z(z) = \langle \Psi_0 |\exp (-{z \cal H}) |\Psi_0 \rangle \equiv \prod_{\bf k} {1+B^2_{\bf k} e^{-2E_{\bf k}({\bf J}_1)z/\hbar} \over 1+B^2_{\bf k} } \ , 
\end{equation}
where 
\begin{equation}
B_{\bf k} = {u_{\bf k}({\bf J}_0)v_{\bf k}({\bf J}_1) - u_{\bf k}({\bf J}_1)v_{\bf k}({\bf J}_0)\over 
u_{\bf k}({\bf J}_0)v_{\bf k}({\bf J}_1) + u_{\bf k}({\bf J}_1)v_{\bf k}({\bf J}_0} \ , 
\end{equation} 
with 
\begin{eqnarray}
&&u_{\bf k}({\bf J}) = \sqrt{{E_{\bf k}({\bf J}) +\epsilon_{\bf k}({\bf J}) \over 2E_{\bf k}({\bf J})}} \ , \nonumber \\
&&v_{\bf k}({\bf J}) = {\rm sign}[\Delta_{\bf k} ({\bf J})] \sqrt{{E_{\bf k}({\bf J}) -\epsilon_{\bf k}({\bf J}) \over 2E_{\bf k}({\bf J})}} \ . 
\end{eqnarray}
The Fisher zeros for the Kitaev honeycomb model (as well as for any BCS-like models) are determined as 
\begin{equation}
z_n({\bf k}) = {\hbar\over 2E_{\bf k}}[\ln (B_{\bf k}^2) +i\pi (2n+1)] \ . 
\end{equation}
Two space dimensions imply that there exist dense areas of Fisher's zeros. These areas cover parts of the real time axis if ${\rm Re} [z_n({\bf k})]=0$ so that $B^2({\bf k}^*)=1$. 
There are intervals on the real time axis
\begin{equation}
T_n^* ={\hbar \pi (2n+1)\over 2E_{{\bf k^*}} } \ , 
\label{Tn}
\end{equation}
which are covered by the areas of zeros. The beginning and the end poits of the intervals $T_n^*$ are determined by the maximum and minimum of Eq.~(\ref{Tn}). If the spectrium of the final Hamiltonian is gapped, the beginning and end points of two consecutive  intervals $T_n^*$ and $T_{n+1}^*$ are equidistant for the values ${\bf k^*}$ minimizing/maximizing $E_{\bf k}$ on the domain $B^2({\bf k}^*)=1$. The length of single intervals increases linearly with $n$. However, if the spectrum of the Kitaev honeycomb model is gapless, those intervals extend to infinity. Notice that 
\begin{equation}
\langle n_{\bf k} \rangle = \sin^2 ({\rm arctan} B_{\bf k}) \ . 
\end{equation}
which implies that for the mode with ${\bf k}^*$ we have $\langle n_{{\bf k}^*}\rangle  =1/2$. The dynamical quantum phase transition imply the change of population from less than 1/2 to the effective populations larger than 1/2 (the inverted occupation). Also, it is important to point out that when quenching to a gapless phase (where there are no zeros in the density of Loschmidt amplitude), excitations cost no energy and, hence, any quench produces inverted mode occupation. If for some
mode ${\bf k}$ its occupation is full $\langle n_{{\bf k}^+}\rangle =1$ (related to the divergence $|B_{{\bf k}^+}| \to \infty$), then it exists necessarily  some  mode $\langle n_{{\bf k}^*} \rangle =1/2$, because $\langle n_{{\bf k}=0} \rangle =0$. It is fulfilled, if the following relation holds 
\begin{equation}
{\epsilon_{{\bf k}^+}({\bf J}_0)\over E_{{\bf k}^+}({\bf J}_0)} = 
- {\epsilon_{{\bf k}^+}({\bf J}_1)\over E_{{\bf k}^+}({\bf J}_1)} = \pm 1 \ . 
\end{equation}
For the quantum quench ending in the gapped phase, e.g., for $J_{x,1} \ge J_{y,1}+J_{z,1}$ and ${\bf k}^+ =(\pi,0)$. Then $\epsilon_{{\bf k}^+} =2(J_{z,1} -J_{x,1}+J_{y,1})$ and $\epsilon_{{\bf k}^+}({\bf J}_1)/E_{{\bf k}^+}({\bf J_1}) =-1$. Both quenches, starting from another gapped phase $J_{y,0} < J_{x,0} +J_{z,0}$, or from the gapless phase $J_{x,0} < J_{y,0} +J_{z,0}$, lead to the non-analytical behavior of the Loschmidt amplitude.  

As we pointed out before, not the Loschmidt echo itself, but rather its time derivative, $l_t$ manifest cusps (dynamical quantum phase transitions) in its time dependence. 

It is also worth to mention the following connection between the characteristics of the Fisher zeros \cite{SK1}
\begin{eqnarray}
&&B^2_{\bf k} = \exp \left( {\pi \hbar (2n+1){\rm Re} [z_n({\bf k})]\over {\rm Im} [z_n({\bf k})]} \right) \ , \nonumber \\ 
&&E_{\bf k} = {(2n+1)\pi \hbar /2{\rm Im}[z_n({\bf k})]} \ . 
\end{eqnarray} 
It implies that the density of Fisher's zeros diverges at the boundary, where ${\bf \nabla} E_{\bf k} \parallel {\bf \nabla} B^2_{\bf k}$. Therefore, the second derivative of ${\bf Re} f(t)$ diverges when approaching the boundary of an interval $T_n^*$ from inside the interval. 

Two important things are also interesting. First, in the one-dimensional limit (if either of $J_{\alpha}$ is zero), the Kitaev model becomes a set of non-interacting between each other spin-1/2 chains. Thus, the critical intervals $T_n^*$ become critical points $t_n^*$, at which the singularities appear in the time dependence of the Loschmidt echo. 

Second, consider the imaginary time axis $z=\tau/\hbar$. It is easy to show that 
\begin{equation}
\lim_{\tau \to \infty} {\cal L}(i\tau) = |\langle 0|\Psi_0\rangle|^4 ={\cal F}^4 \ , 
\end{equation}
where $|0\rangle$ is the eigenfunction of the ground state energy state of the final Hamiltonian, and ${\cal F}$ is the fidelity, i.e., the Loschmidt echo is related to the fidelity in the large imaginary time limit. For the Kitaev honeycomb model the fidelity is 
\begin{equation}
{\cal F} = |\langle 0|\Psi_0\rangle |= \exp \left(- {L\over 8\pi^2} \int d^2k \ln [1+B_{\bf k}^2] \right) \ . 
\end{equation}
Hence, it is the direct way of observing the fidelity after quantum quenches, to study the density of the return probability at long time scales. 

Summarizing, in two-dimensional systems Fisher's zeros are organized to areas rather than lines for one-dimensional systems. The covering of intervals of real time axis by such areas shows critical points in the time evolution of the Loschmidt amplitude. This leads to dynamical quantum phase transitions, as discontinuities in the second time derivative of the density of the return probability (dynamical free energy), rather than discontinuities of the first derivative for one-dimensional quantum systems. 

\section{Non-integrable models}

So far dynamical quantum phase transitions were considered in only integrable models. It is interesting to understand how the non-integrability can affect the existence and features of dynamical quantum phase transitions. 

Several studies considered the behavior of the Loschmidt amplitude after quantum quenches across quantum critical points for non-integrable systems numerically and analytically. For example, Ref.~\cite{AS}  studied the quantum quenches for the XXZ spin-1/2 chain in the external homogeneous and inhomogeneous (staggered) field. The latter yields non-integrabiliity. Using the light-cone renormalization group, it was shown \cite{AS} that quenches of the staggered field value $h_{st}$ across the quantum critical point $h_{st}=0$ can produce features in the behavior of the logarithm of the next-order eigenvalues of the quantum transfer matrix $-\ln|\Lambda_n|^2$, which is responsible for the correlation lengths. Though the density of the return probability  does not show cusps in the time dependence for such quenches at small times, but do show cusps for longer times. It is related to the fact that lines of Fisher's zeros cross the time axis at larger times, and do not cross the time axis at small, times (though approaching time axis). 

The paper Ref.~\cite{KS} has studied dynamical quantum phase in the transverse field Ising chain with additional interactions, which remove the exact integrability, numerically, using the time-dependent density matrix renormalization group approach. The authors have studied the axial next-nearest-neighbor Ising model (ANNNI) \cite{ANNNI} with the Hamiltonian 
\begin{equation}
{\cal H}_{ANNNI} = - \sum_j [2J(S_j^zS_{j+1}^z +\Delta S_j^zS_{j+2}^z) + H\sigma_j^x] \ ,
\end{equation}
where $J >0$. For $\Delta =0$ the transverse field Ising model is recovered. The ground state phase diagram of the ANNNI model has four phases \cite{phase}, namely, the paramegnetic phase, the ferromagnetic phase with doubly degenerate ground state, the so-called "antiphase" with the doubling of the N\'eel-like structure along the chain, and the "floating" phase between the paramagnetic phase and the "antiphase". The phase transition between the ferromagnetic phase and the paramagnetic one belongs to the Ising class of universality with the critical exponent $\nu=1$. For $\Delta <0$ the critical boundary is determined by the equation ($h =H/2J$)
\begin{equation}
1+2\Delta = h_c + {\Delta h_c^2\over 2((1+\Delta)} \ . 
\end{equation}
Numerical calculations manifest kinks in the time behavior of the density of the return probability and features for the time dependence of the order parameter for the quantum quenches across critical lines, i.e. the dynamical quantum phase transitions. 

The ANNNI model was also studied numerically and analytically in \cite{KKK} using the continuous unitary transformation approach \cite{CUT,MK}. In that approach the Hamiltonian gets a diagonal form after the series of infinitesimal unitary transformations. The evolution of the Hamiltonian under that set is parametrized by a flow parameter $l$. It is governed by the equation
\begin{equation}
{\cal H}_l (l) =[\Gamma (l),{\cal H}(l)] \ , 
\end{equation}
where $A_l \equiv d A/dl$, and $\Gamma(l)$ is an antihermitian generator. If the Hamiltonian before the quench is ${\cal H}_0$ and after quench is ${\cal H} ={\cal H}_0 +{\cal H}_1$, then ${\cal H}(l=0) ={\cal H}$ is the initial condition. At finite $l$ we transform ${\cal H}(l) =U(l){\cal H}(0)U^{\dagger}(l)$, where $U_l(l)=\Gamma(l)U(l)$, and $U(0)=I$ (the unity matrix). The flow converges to the fixed point ${\cal H}(\infty)$ ("energy diagonal"), if the generator $\Gamma(l)$ is chosen appropriately. For the fixed point Hamiltonian we can chose ${\cal H}_0$, and set $\Gamma (l) =[{\cal H}_0,{\cal H}_1]$, which produces the fixed point at which $[{\cal H}_0,{\cal H}(\infty)] =0$.  For the quantum quench for the ANNNI model this approach yields renormalization of density of the return probability (the Loschmidt echo) due to nonzero $\Delta$
\begin{eqnarray}
&&l(\Delta,t) = -{1\over \pi} \int_0^{\pi} dk \ln [\cos^2\phi_k 
\nonumber \\ 
&&+ \sin^2 \phi_k e^{-2it{\tilde \varepsilon}_k/\hbar} ] - l^{(1)}(t) + \dots
\end{eqnarray}
where ${\tilde \varepsilon}_k$ and $l^{(1)}(t)$ are the renormalized energy of the excitiations of the final Hamiltonian, and the (real) correction to the return probability due to nonzero $\Delta$, respectively \cite{KKK}. For small $\Delta$ the value $l^{(1)}(t)$ is small. However, secular terms are introduced due to the truncation procedure in the continuous unitary transformation approach. The comparison of the results of this analytic approach with the ones of the numerical calculation using the time-dependent desity matrix renormalization group \cite{KKK} shows very good agreement between them. Hence, the analytic approach for the dynamical quantum phase transitions in non-integrable models also reveals existence of the former.  

The second non-integrable model, studied numerically in \cite{KS} (see also \cite{SSD}), is the Ising chain in the tilted magnetic field. Also, the dynamical quantum phase transitions manifest itself as non-analyticities in the time dependence of the density of the return probability (the Loschmidt echo). 

These studies show, that the appearance of the dynamical quantum phase transitions is not an artifact of the integrable models. The non-analyticities in the time dependence of the Loschmidt echo (the dynamical free energy) are stable with respect to inclusion of  integrability-breaking perturbations. 
 
\section{Topology and dynamical quantum phase transitions}

We can generalize the results for dynamical quantum phase transitions for the general case of gapped two-band fermionic Bogolyubov-de Gennes models \cite{VD2,BH}, restricting our consideration to the space dimensions  one and two, where quantum phase transitions are expected to take place in equilibrium. We can consider the Hamiltonian of this class of models using the Nambu representation
\begin{equation}
{\cal H}   = \sum_{\bf k} {\bf c}_{\bf k}^{\dagger} H_{\bf k} {\bf c}_{\bf k} \ , 
\end{equation} 
where ${\bf c}_{\bf k}^{\dagger} = (c_{\bf k}^{\dagger}, c_{-{\bf k}})$ is  the spinor, and 
\begin{equation}
H_{\bf k} = ( {\bf d}_{\bf k} \cdot {\bf \sigma}) \ , 
\end{equation}
where ${\bf \sigma}$ is the Nambu pseudospin. Obviously, the particle-hole symmetry implies 
$\sigma^x H_{\bf k} \sigma^x = -H^*_{-{\bf k}}$. Then, it is clear that $d^{x,y}_{\bf k}$ have to be odd functions of ${\bf k}$, and $d^z_{\bf k}$ has to be an even function of ${\bf k}$. In the space dimension one for $k=0,\pi$ it implies that $H_{k=0,\pi}$ has only $z$-component in the Nambu representation, because $d_{k=0,\pi}^{x,y}$ have to vanish. Two topologically different classes of such Hamiltonians differ from each other by the sign of $d^z_{k=0}d^z_{k=\pi}$, which is negative in the topologically nontrivial superconductor \cite{Kit}. 

The quantum quench, i.e., the sudden change of the parameters of the considered Hamiltonian implies ${\bf d}_{\bf k}(t=0) ={\bf d}_{\bf k}^0$ and ${\bf d}_{\bf k}(t>0) ={\bf d}_{\bf k}^1$.The density of the Loschmidt amplitude can be written as \cite{VD2}
\begin{equation}
g(t) =-{1\over L} \sum_{\bf k} \ln \biggl[ \cos (\varepsilon_{\bf k}^1t/\hbar) +i {({\bf d}_{\bf k}^{1}\cdot {\bf d}_{\bf k}^0)\over \varepsilon_{\bf k}^0\varepsilon_{\bf k}^1} \sin (\varepsilon_{\bf k}^1t/\hbar)\biggr] \ , 
\end{equation}
where $\varepsilon_{\bf k}^{0,1}  = |{\bf d}_{\bf k}^{0,1}|$. Fisher's zeros are 
\begin{equation}
z_n({\bf k})= {i\pi \hbar (2n+1)\over 2\varepsilon_{\bf k}^1} - {\hbar \over \varepsilon_{\bf k}^1}
\arctan \left( {({\bf d}_{\bf k}^{1}\cdot {\bf d}_{\bf k}^0)\over \varepsilon_{\bf k}^0\varepsilon_{\bf k}^1}\right) \ .  
\end{equation}
It is clear that the lines of Fisher's zeros approach the imaginary axis when $({\bf d}^0_{{\bf k}^*}\cdot 
{\bf d}^1_{{\bf k}^*})=0$ at ${\bf k} ={\bf k^*}$. The states of the Hamiltonian in the Nambu representation are in one-to-one correspondence with the states of a vector ${\bf d}_{\bf k}$ on the Bloch sphere of the radius $\varepsilon_{\bf k}$. Then, at Fisher's zeroes the vector of the initial state is perpendicular to the vector of the finite state. It is connected with the fact that dynamical quantum phase transitions are related to the initial state orthogonal to the time-evolved (final) state. Such a condition relates dynamical phase transitions to the topology of the initial and final Bogolyubov-de Gennes superconductors \cite{VD2}. We can get similar conclusion using a little different arguments. Namely, let the system to have initially occupied lower Bloch states with the wave function $|u^{0-}_{\bf k} \rangle$. Then the time-dependent (finite) wave function for each ${\bf k}$ can be written as 
\begin{equation}
|\psi_{\bf k} (t)\rangle = e^{i\varepsilon_{\bf k}^1 t/\hbar}g_{\bf k} |u_{\bf k}^{1-}\rangle + 
e^{-i\varepsilon_{\bf k}^1 t/\hbar}e_{\bf k} |u_{\bf k}^{1+}\rangle \ , 
\end{equation}
where $|u_{\bf k}^{1\pm}\rangle$ are Bloch states of the final Hamiltonian (with energies $\pm \varepsilon_{\bf k}^1$), and 
\begin{eqnarray}
&&|g_{\bf k}|^2 = \langle u_{\bf k}^{1-}|u_{\bf k}^{0-}\rangle = = {1\over 2} \left[ 1+ {({\bf d}_{\bf k}^{1}\cdot {\bf d}_{\bf k}^0)\over \varepsilon_{\bf k}^0\varepsilon_{\bf k}^1} \right] \ , 
\nonumber \\ 
&&|e_{\bf k}|^2 = \langle u_{\bf k}^{1+}|u_{\bf k}^{0-}\rangle = = {1\over 2} \left[ 1- {({\bf d}_{\bf k}^{1}\cdot {\bf d}_{\bf k}^0)\over \varepsilon_{\bf k}^0\varepsilon_{\bf k}^1} \right] \ . 
\end{eqnarray}
At $t=t_n^*$ we have $|g_{{\bf k}^*}|^2 = |e_{{\bf k}^*}|^2$, i.e. dynamical quantum phase transitions occur when initial lower Bloch state is an equal weight superposition of the final Bloch states \cite{BH}. 

Topological superconductors with the Bogolyubov-de Gennes Hamiltonians are important for the construction of topological quantum computers \cite{topqc}, where edge Majorana states play the principal role \cite{Major}. Such models were realized in recent experiments following theoretical predictions \cite{Majexp,ferr}.

There are several aspects of the connection of dynamical quantum phase transitions with the topology. First, the dynamical topological order parameter was introduced \cite{BH}. Namely, let us consider the Loschmidt amplitude $G(t) = \prod_{{\bf k}>0} G_{\bf k}(t)$, and introduce polar co-ordinates so that $G_{\bf k}(t) = r_{\bf k}(t) \exp (i \varphi_{\bf k}(t))$. The phase $\varphi_{\bf k}(t)$ has two contributions:
\begin{equation}
\varphi_{\bf k}(t) = \varphi_{\bf k}^G(t) +\varphi_{\bf k}^{dyn}(t) \ , 
\end{equation}
where the dynamical phase is 
\begin{equation}
\varphi_{\bf k}^{dyn} (t) = -\int_0^{t/\hbar} \langle \psi_{\bf k}(s)|{\cal H}|\psi_{\bf k}(s)\rangle = \varepsilon_{\bf k}^1 t(|g_{\bf k}|^2 -e_{\bf k}|^2)/\hbar  \ . 
\end{equation} 
As for $\varphi_{\bf k}^G(t)$, it is the Berry phase. In one space dimension for $k=0,\pi$ either $|g_k|^2=0$ or $|e_k|^2=0$, and $\varphi_{k=0,\pi}(t)= \varphi_{k=0,\pi}^{dyn}(t)$. It means that the Berry phase is equal to zero at $k=0,\pi$. Hence, the interval $0 < k < \pi$ can be endowed with the topology of the unit circle $S^1$ by identifying its end points. Such a periodic structure can be considered as the effective Brillouin zone. Then, the dynamical topological order parameter can be defined as 
\begin{equation}
\nu_D(t) = {1\over 2\pi} \oint_0^{\pi} {\partial \varphi_k^G(t)\over \partial k} dk \ . 
\end{equation}
One can see that $\nu_D(t)$ has integer values. It is the winding number of the Berry phase over the effective Brillouin zone. Without Fisher's zeros it smoothly depends on time. However, it jumps from one integer to the other at $t=t_n^*$. It is constant in the time intervals between $t^*_n$. If for $k=0,\pi$ the vector ${\bf d}_k$ is directed to the north or south poles of the Bloch sphere, then critical momenta are located at the equator of the latter. Then the change of the dynamical topological order parameter is related to whether ${\bf d}^1_k$ traverses the equator of the Bloch sphere from the northern to the southern hemisphere $\delta \nu_D(t_n^*) = {\rm sgn} (s_{k^*}) =-1$ (where $s_{k^*} =(\partial |e_{k}|^2/\partial k)|_{k=k^*}$), and $\delta \nu_D(t_n^*) = 1$ for traverses from the southern to the northern pole. Formally, $\nu_D(t)$ is the topological invariant distinguishing homotopically non-equivalent mappings of the effective Brillouin zone to $U(1),k \to \exp [i \phi_k^G(t)]$ from the unit circle $S^1$ to itself. Jumps of $\nu_D(t)$ can be negative (for transitions between topologically non-equivalent equilibrium states), or alternating negative and positive (the latter is characteristic for systems with two kinds of Fisher's zeros, which happens for transitions between topologically equivalent equilibrium phases) \cite{BH}. 

The dynamical topological order parameter distinguishes periods of time evolution, which are separated by dynamical quantum phase transitions. It is different from standard topological invariants \cite{topinv}, because it is dynamical in nature. dynamical quantum phase transitions can happen between topologically different or topologically equivalent equilibrium states. 
 
Let us consider then one-dimensional and two-dimensional topological insulators, starting with 
the one-dimensional  ones of the chiral (AIII symmetry class) and chiral and particle-hole symmetry (BDI class) \cite{topinv}. For these classes ${\bf d}_{k}$ lie in the $xy$ plane. The topological number is the winding number, i.e., the number of times vector ${\bf _d}_k$ winds around the origin when $k$ sweeps through the Brillouin zone \cite{VD2}
\begin{equation}
\nu = {1\over 2\pi} \int dk ({\hat {\bf d}}_k^x \partial_k {\hat{ \bf d}}_k^y - {\hat {\bf d}}_k^y \partial_k
 {\hat {\bf d}}_k^x ) \ , 
\end{equation}
where ${\hat {\bf d}}_k \equiv {\bf d}_k/|{\bf d}_k|$, and $\partial_k = \partial /\partial k$. The authors of Ref.~\cite{VD2} have formulated the following statement. If the winding number of two vector fields ${\bf d}_k^0$ and ${\bf d}_k^1$ defined on the Brillouin zone $S^1$ differ by the integer $\Delta \nu$, 
the image of the scalar product $({\hat {\bf d}}_k^0\cdot {\hat {\bf d}}_k^1)$ covers the interval $[-1,1]$ at least $2\Delta \nu$ times. It implies that Fisher's zeros sweep through the real axis $2\Delta \nu$ times while $k$ goes through the Brillouin zone. Hence, there are at least $2\Delta \nu$ points in the $k$-space, were ${\bf d}_k^0$ and ${\bf d}_k^1$ are perpendicular (the condition of dynamical quantum phase transitions. If the ground state winding number of the initial (final) state of the topological insulator is $\nu_0$ ($\nu_1$), then the angle of rotation of ${\bf d}_k^{i}$ is a smooth function, which differs by $2\pi \nu_i$ at $k=-\pi$ and $\pi$ for $i=0,1$. The change of the angle of rotation $\Delta \phi_k$ between the initial and final states is $2\pi \delta \nu$, and, therefore, 
$({\hat {\bf d}}_k^0\cdot  {\hat {\bf d}}_k^1) =\cos (\Delta \phi_k)$ covers the interval $[-1,1]$ at least $2\Delta   \nu$ times. The line of Fisher's zeros can be doubly  degenerate if further symmetries (like inversion or time-reversal ones) connect states with $k$ and $-k$, and, hence, only $\Delta \nu$ non-equilibrium time scales can exist (see above). 

If the time-reversal symmetry is broken (D symmetry class \cite{topinv}) ${\bf d}_k$ can have $z$ components. The $Z_2$ invariant is 0 (topologically trivial) if ${\hat {\bf d}}_{k=0} ={\hat {\bf d}}_{k=\pi} = (0,0 \pm 1)$, and it is 1 (topologically nontrivial) if  ${\hat {\bf d}}_{k=0} =-{\hat {\bf d}}_{k=\pi}$. If the quantum quench connects phases with different topology, e.g., $\nu_o=1$ and $\nu_1=0$, then there must be a quasimomentum $k^*$ for which the vectors are orthogonal, $({\bf d}_{k^*}^0\cdot {\bf d}_{k^*}^1)=0$, hence $({\hat {\bf d}}^0_{k}\cdot {\hat {\bf d}}_k^1)$ covers the interval $[-1,1]$. 

It turns out that ${\bf d}_{\bf k}^0$ and ${\bf d}_{\bf k}^1$ can be orthogonal accidentally. 

In two space dimensions the Chern number ${\cal C}$ is the topological number for two-band topological insulators 
\begin{equation}
{\cal C} = {1\over 4\pi} \int_{BZ} dk_xdk_y ({\hat {\bf d}}_{\bf k}\cdot [\partial_{k_x} {\hat {\bf d}}_{\bf k} \times \partial_{k_y} {\hat {\bf d}}_{\bf k}]) \ . 
\end{equation}
The Chern number counts how many times the surface defined by ${\hat {\bf d}}_{\bf k}$ covers the unit sphere. It is possible to show that if the Chern numbers of two vectors ${\bf d}_{\bf k}^0$ and ${\bf d}_{\bf k}^1$ on the Brillouin zone $T^2$ differ in modulus ${\cal C}_1 \ne {\cal C}_2$ (except of ${\cal C}_1 =-{\cal C}_2$), then the image of the scalar product 
${\hat {\bf d}}^0_{\bf k} \cdot {\hat {\bf d}}_{\bf k}^1$ is $[-1,1]$. and dynamical quantum phase transitions necessarily occur \cite{VD2}. 

Hence, Fisher's zeros connect $\pm \infty$ if the modulus of Chern's number is changed under the quench (though there can be such a connection for the same Chern's numbers). 

For the superconductor $k$ is taken for the half of the Brillouin zone. However, the particle-hole symmetry causes that exactly the same contribution comes from the other half of the Brillouin zone. Hence, the Loschmidt amplitude can be formulated on the total Brillouin zone. Hence, for quantum quenches connecting superconducting phases with different modules of Chern's numbers, dynamical quantum phase transitions have to exist. 

It is also important to connect dynamical quantum phase transitions with the entanglement. The latter measures the time evolution of wave functions without reference to any observables \cite{Ch}.  The evolution of the entanglement manifests qualitative differences for quantum quenches where $({\bf d}_{\bf k}^0 \cdot {\bf d}_{\bf k}^1)=0$ for some ${\bf k}$, i.e., the condition for the existence of dynamical quantum phase transitions. 

As an example, consider following \cite{VD2} the Haldane model \cite{Hal}, which is the model of electrons which hop on a honeycomb lattice between the next-nearest-neighbor sites with an artificial magnetic field. For this model 
\begin{equation}
{\bf d}_{\bf k} = ({\rm Re} \{ f(k)\},{\rm Im} \{f(k)\}, m-g({\bf k},\varphi)) \ , 
\end{equation}
where $f(k) =\gamma_1 \sum_j \exp (-i{\bf k}{\bf \delta}_j)$, the vectors ${\bf \delta}_j$ connect three nearest neighbors of the honeycomb lattice ($\gamma_1$ is the hopping amplitude between nearest neighbors), $m$ is the mass (which describes the homogeneous staggered lattice potential), and $g({\bf k},\varphi)$ describes the nextnearest-neighbor hopping (due to the staggered magnetic field). The latter yields a nontrivial topology. The Chern number depends on the phase $\varphi$, the next-nearest hopping amplitude $\gamma_2$ , and $m$. It is zero, ${\cal C} =0$ if 
$|m| > |3\sqrt{3} \gamma_2 \sin \varphi |$, and ${\cal C }= \pm 1$ if 
$m <  |3\sqrt{3} \gamma_2 \sin \varphi |$. The sign depends on $\varphi$ and $\gamma_2$. 

For two-dimensional systems Fisher's zeros form areas rather than lines in the space dimension one. Each Fisher's area corresponds to $n$ of $z_n({\bf k})$. It is parametrized by two values $k_x$ and $k_y$. If a Fisher area crosses the imaginary axis, the density of the Loschmidt amplitude manifest cusps at boundaries of Fisher's area, i.e., the second derivative of the density of the Loschmidt amplitude jumps at the boundary of Fisher's area. The size of the jump is proportional to the density of Fisher's zeros normalized by the system size. If the density of Fisher's zeros diverges, then the slopes of cusps in ${\rm Re} f'(t)$ inside the Fisher area diverge similarly. It occurs  not only in the Haldane model, but also in the so-called BHS model \cite{BHS}, or in the lattice version of the chiral $(p +ip)$-superconductor \cite{sup}, in which ${\bf d}_{\bf k} = (A \sin k_x, A\sin k_y, \Delta + \cos k_k +\cos k_y)$, where $A$ is the hopping amplitude, and $\Delta$ is the pairing amplitude. 

Dynamical quantum phase transitions are related to the existence of Fisher's zeros or nodes in the wavefunction overlap (Loschmidt echo) between the initial state and eigenstates of the post-quenched Hamiltonian. These nodes are topologically protected if participating wave functions have distinctive topological indexes. The condition of the existence of dynamical quantum phase transitions $G(t) =0$ can be interpreted geometrically as Fisher's zeros $z_n(t) =|\langle n \Psi_0\rangle|^2 \exp (-itE_n/\hbar)$ form a closed  polygon in the complex plane at $t=t^*$ \cite{HB}. Hence, amplitudes of Fisher's zeros satisfy a triangle inequality $\sum_{m\ne n} |z_m| \ge |z_n|$. Taking into account that $\sum_n |z_n| =1$, one obtains \cite{HB}
\begin{equation}
|z_n| = |\langle n |\Psi_0\rangle|^2 \le {1\over 2} \ . 
\label{ac}
\end{equation}
On the other hand, the dynamical phases of Fisher's zeroes form a subspace $M$ on the $N$-torus ($N$ is the total number of Fisher's zeros). Let $G(t) =\prod_{\bf k} G_{\bf k}(t)$. As long as the gaps $E_{{\bf k},n+1} -E_{{\bf k},n}$ are not rationally related, the dynamical phases are ergodic on the $N_{\bf k}$-torus ($N_{\bf k} -1$ is the number of such gaps) and will evolve into its subspace $M$. 
If the phase ergodicity holds for all ${\bf k}$ (i.e. there is no degeneracy at any ${\bf k}$) then dynamical quantum phase transitions imply that at least one ${\bf k}$ exists, for which the amplitude condition Eq.~(\ref{ac}) is satisfied for such ${\bf k}$ mode. Hence dynamical quantum phase transitions arize from the nodes in the wave function overlaps \cite{HB}. In the space dimension one, the Berry phase of a real Bloch band is quantized to $0$ or $\pi$. The overlap of two bands with different Berry phases must have at least one node. On the other hand, in the space dimension two, the overlap of Bloch bands with the Chern numbers ${\cal C}_1$ and ${\cal C}_2$ must have at least $|{\cal C}_1 -{\cal C}_2|$ nodes in the Brillouin zone \cite{HB}. It means that nodes in the wave function overlaps are topologically protected if the topological characteristics (the Berry phase, or the Chern number) are different. 

\section{Broken symmetry in dynamical quantum phase transitions}

We will illustrate the broken symmetry in dynamical quantum phase transitions using as the main example the XXZ chain following \cite{Hey1}. 

Consider the XXZ spin-1/2 chain with the Hamiltonian Eq.~(\ref{XXZ}) in zero magnetic field. For the antiferromagnetic case $\Delta >0$ there exists a quantum phase transition at $\Delta=1$ of the Berezinskii-Kosterlitz-Thouless type between the Luttinger liquid phase and the gapped antiferromagnetic phase. The order parameter of such a transition is the staggered magnetization 
\begin{equation}
\langle M_s \rangle = {1\over L} \sum_j (-1)^j \langle S_j^z \rangle \ . 
\end{equation}
Suppose in the quantum quench the system was initially in the N\'eel state $|\Psi_0\rangle = |\uparrow \downarrow \rangle = | \uparrow \downarrow \uparrow \downarrow \dots \rangle$. This state is degenerate with $|\downarrow \uparrow \rangle$. It is equivalent  to the state of the system at $\Delta_0 \to  \infty$. In Ref.~\cite{Hey1} using exact diagonalization based on the Lanczos tri-diagonalization of the Hamiltonian with full re-orthogonalization \cite{CW} were performed.  The order parameter as a function of time is monotonic for large time scales, and it is oscillatory at small times (of order of $\hbar/J$) as $\Delta$ crosses the equilibrium quantum critical point. If the initial Hamiltonian commutes with the order parameter at least at one point of the parameter space (here $\Delta_0 \to \infty$), then the energy and staggered magnetization can be measured simultaneously. Then it is possible to decompose the operator of the order parameter spectrally during the dynamical evolution 
\begin{equation}
\langle M_s \rangle = \int d \varepsilon M_s(\varepsilon,t) P(\varepsilon,t) \ , 
\label{Ms}
\end{equation}
where $P(\varepsilon, t)$ is the probability distribution that the system has the energy density $\varepsilon$ at time $t$, 
\begin{equation}
P(\varepsilon,t) =\sum_n |\langle n|\Psi_0(t) \rangle|^2 \delta (E_n/L -\varepsilon) \ , 
\end{equation}
with $|\Psi_0(t) \rangle =\exp (-i{\cal H} t/\hbar) |\Psi_0 \rangle$. $M_s(\varepsilon,t)$ is the contribution to the full expectation value of the (time-dependent) order parameter from the energy density. Energies are measured with the initial, not the final Hamiltonian. Thereby, the "exclusive" perspective \cite{CTH} is chosen in which the perturbation which generates the dynamics is not included into the internal energy of the system. 

Due to the twofold degeneracy of the initial ground state we can write
\begin{equation}
P(0,t) = {\cal L}_{\uparrow \downarrow} (t) +  {\cal L}_{\downarrow \uparrow} (t) \ , 
\end{equation}
where ${\cal L}_{\eta} = |\langle \eta |\psi_0(t)\rangle |^2$ with $\eta = |\uparrow \downarrow\rangle$ or $|\downarrow \uparrow \rangle$. In the thermodynamic limit $L \to \infty$ each of microscopic probabilities ${\cal L}_{\eta} (t) =\exp [-L\lambda_{\eta}(t)]$, i.e., it obeys the large deviation scaling \cite{Tou} with $\lambda_{\eta}(t)$ intensive \cite{GS,S}. Hence, one of the probabilities always dominate $P(0,t) = \exp [-L\lambda(t)]$, where $\lambda(t) = {\rm min}_{\eta} \lambda_{\eta}(t)$ within the exponential accuracy. It is obvious, taking in account the definition of the Loschmidt amplitude, that ${\cal L} (t) = |G(t)|^2$, i.e. the return probability.   Hence, it has to manifest dynamical quantum phase transitions. Really, Ref.~\cite{Hey1} shows using exact diagonalization that such a transition really occurs in the XXZ chain as a kink it time dependence of $\lambda(t)$ due to the crossover of $\lambda_{\eta}(t)$ at some value $t=t^*$. It is possible to detect the dynamical quantum phase transition (which exists only in the thermodynamic limit) from the finite size calculations with the high accuracy. 

Due to mentioned commutation of the operator of the order parameter and the Hamiltonian, we can write 
\begin{equation}
\langle M_s \rangle = \int d \varepsilon \int d m_s m_sP(\varepsilon, m_s, t) \  , 
\end{equation}
where $P(\varepsilon,m_s,t)$ is the joint distribution function that the system has the energy density $\varepsilon$ and staggered magnetization density $m_s$ at time $t$. First one measures the eigenstate with the energy density $\varepsilon$, followed by the measurements of the staggered magnetization. In the thermodynamic limit the distribution $P(\varepsilon,m_s,t)$ satisfies the central limit theorem, i.e., at given $\varepsilon$ only a narrow region (it is vanishing in the thermodynamic limit) mainly contributes in the vicinity of $m_s = M_s (\varepsilon,t)$, where $P(\varepsilon,m_s,t)$ is maximal. Then we get  Eq.~(\ref{Ms}), which implies $P(\varepsilon,t) = \int m_s P(\varepsilon, m_s,t)$. Then, according to \cite{Tou} it is possible to calculate 
\begin{equation}
M_s (\varepsilon,t) =\langle \Psi_0 (t,s) M_s \Psi_o (t,s)\rangle 
\end{equation}
where ${\cal N}^{-1/2}|\Psi_0(t,s) \rangle = \exp( -{\cal H}_0 s/2 |\Psi_0 (t) \rangle$, as in the full counting statistics approach, cf. above. Here ${\cal N} (s,t)= \langle \Psi_0(t)| \exp(-{\cal H}_0 s) |\Psi_0(t) \rangle$, and $s(\varepsilon,t)$ is the solution of the equation
\begin{equation}
\varepsilon = L^{-1} \langle \Psi_0 (s,t) |{\cal H}_0 |\Psi_0(s,t)\rangle \ . 
\end{equation}
Using numerical calculations Ref.~\cite{Hey1} has shown that the nonanalyticity of the density of probability (dynamical quantum phase transition) translates into the non-analyticity of the zero energy limit of the order parameter $\langle M_s(0,t) \rangle$. In the thermodynamic limit the main contribution to $\langle M_s(t) \rangle$ comes from the narrow interval in the vicinity of $\varepsilon_{av} = L^{-1}\langle {\cal H}_0(t)\rangle$ due to the central limit theorem. Then $\langle M_s(t) \rangle \to M_s(\varepsilon_{av},t)$ Then one can show \cite{Hey1} that $M_s(0,t) \to M_s(\varepsilon_{av}t)$, demonstrating that dynamical quantum phase transitions control the oscillatory decay of the order parameter. 

These results can be generalized to other models, if the initial Hamiltonian exhibits the ground state degeneracy, and if the initial Hamiltonian has one point in the parameter space, where it commutes with the order parameter. It can be related to Ising models with the vanishing transverse field, to the Hubbard systems (both Fermi and Bose) at vanishing hopping in the charge-density wave limit, does not matter what the dimension of space is.

\section{Scaling and universality for dynamical quantum phase transitions}

Consider, following \cite{Hey2} the two-dimensional transverse field Ising model ($\sigma^z_j \equiv 2S^z_j$) 
\begin{equation}
{\cal H} = -\sum_{l,m} J_{lm} \sigma^z_l \sigma_m^z -H \sum_l  \sigma_l^x \ . 
\end{equation}
The nearest neighbor coupling constants $J_{lm} >0$ are taken such that $J_{lm} =J$ along rows and $J_{lm} =J_{\perp}$ along columns. We have shown in the previous sections that there exist dynamical quantum phase transitions in the transverse field Ising model for the space dimentions one and two. 

The initial Hamiltonian can be considered for $H \to \infty$, i.e., the ground state wave function is the superposition of all spin configurations in the $\sigma_m^z$ basis. Then at $t=0$ the quantum quench takes place to the state with $H=0$. Then the Loschmidt amplitude can be written as 
\begin{equation}
G(t) = {1\over 2^L} {\rm Tr} \left( e^{(it/\hbar)\sum_{l,m} J_{lm}\sigma_l^z\sigma_m^z}\right) \ . 
\end{equation}  
Let us introduce the notations $K=iJt/\hbar$ and $K_{\perp} =itJ_{\perp}/\hbar$. 
Consider first the case $J_{\perp} =0$. In this case the Ising model can be solved using the transfer matrix method $G(t) ={\bf Tr} {\hat T}^L$, where the eigenvalues of the transfer matrix $\hat T$ are $\nu_c =\cosh K$ and $\nu_s = \sinh K$ \cite{Zvbook1}. In the thermodynamic limit $L \to \infty$ the Loschmidt amplitude is dominated by the largest eigenvalue $G(t) = \nu^L$, with $\nu = \nu_c$ if $|\nu_c| > |\nu_s|$, and $\nu =\nu_s$ in the opposite case.  Obviously, $\nu$ is switched from $\nu_c$ to $\nu_s$, producing non-analytisity of the density of the return probability $l(t) = -2 {\rm Re} [\ln (\nu)]$. The critical times at which dynamical quantum phase transitions take place, are $t_n^* =\pi \hbar (2n+1)/4J$. In equilibrium (with $it/\hbar \to 1/T$ the condition $|\nu_c|=|\nu_s|$ can be realized only in the ground state limit $T=0$. 

Let us realize the renormalization group scheme for the complex parameter space \cite{RG}. For this we eliminate every second spin via decimation. The partition function $Z(K,L)= {\rm Tr} [\exp ({\cal H}(K,L))]$, where $L$ is the size of the chain, can be written as $Z(K,L) = Z(K',L/2) = {\rm Tr} [\exp ({\cal H} (K', L/2))]$ with the renormalized value of the effective coupling $K'$ for remaining $L/2$ spins (periodic boundary conditions used). Pauli matrices commute, then using the equation 
$\exp K \sigma^z (\sigma^z)' = \cosh K + \sigma^z (\sigma^z)' \sinh K$ one gets 
\begin{equation}
Z(K,L) = {\rm Tr} \prod_{l=0}^{L/2}  \cosh^2 K  (1+ \tanh^2 K) \sigma^z_{2l}\sigma^z_{2l+2}) \ .
\end{equation} 
Then the exact renormalization group transformation can be written as (cf. \cite{Kubo})   
\begin{equation}
\tanh (K') = \tanh^2 (K)  \ ,\ e^{2K'} = \cosh (2K ).  
\end{equation}
It follows that the renormalization group has two fixed points: $K^*=0$ and $K^* =\infty$, related to the equilibrium ones at $T = \infty$ and $T=0$, even if $K$ is complex. For small $|K| \ll 1$ we get $K' =K^2$, implying the fixed point $K^* =0$ is stable. Then for $K =K^* +\delta K$ in the vicinity of the other fixed point $K^* = \infty$ we get $\delta K' = 2\delta K \equiv b^{\lambda} \delta K$, which means $b=2$ and $\lambda = 1$, for the associated change of the length scale and anomalous dimension after decimation, respectively. Hence, the fixed point $K^* = \infty$ is unstable, as in equilibrium. However, if we consider not the linear regime for the initial coupling, it can be stable. The dynamical quantum phase transitions at times $t^*_n$ map onto the fixed point $K^* =\infty$ after two renormalization group steps. Consider the weak deviation $\tau = (t-t_c)/t_c$ from the dynamical quantum phase transition. Then using $\lambda =1$ times with weak deviation from $t_c$ map after two renormalization group onto the linear regime of the unstable fixed point, yielding \cite{Hey2} the scaling form 
\begin{equation}
g(\tau) \propto |\tau|^{d/\lambda} \Phi_{\pm} \, 
\end{equation}
where $\Phi_{\pm}$ are constants, which can be different for $\tau >0$ and $\tau <0$. Therefore, for dynamical quantum phase transitions one can attribute this scaling behavior at the unstable fixed point of the renormalization group, extending the scaling, and universality to the non-equilibrium regime. 

Notice, that despite the special quantum quench was considered, the identificaltion of dynamical quantum phase transitions with unstable fixed points allow the conclusion that the weak symmetry-preserving perturbations do not change the above mentioned universality. Let us demonstrate it for nonzero transverse fields $H \ll J$. The field part using the standard time-dependent perturbation theory can be eliminated.  Consider the transverse field Ising chain with the weak field $H$ as the perturbation $V$ and the Ising part as the main Hamiltonian ${\cal H}_0$. Then $\exp (-i{\cal H} t/\hbar) = \exp(-i{\cal H}_0t/\hbar)W(t)$, where 
\begin{eqnarray}
&&W(t) ={\cal T} e^{-i\int_0^tdt' V(t')/\hbar} \ , \nonumber \\ 
&&V(t) = \exp (i {\cal H}_0t/\hbar) V \exp(-i {\cal H}_0t/\hbar) \ ,  
\end{eqnarray}
where $\cal T$ denotes the time ordering. Then using the standard cumulant expansion \cite{Hey2} 
one gets $G(t) = 2^{-L}{\rm Tr} \exp ({\cal H}_{eff})$, where 
\begin{equation}
{\cal H}_{eff} = {\tilde K}\sum_l\sigma_l^z\sigma_{l+1}^z + G\sum_l\sigma_l^z\sigma_{l+2}^z \ , 
\end{equation} 
with ${\tilde K} = itJ/\hbar +H[1-\cos(4Jt/\hbar)]/4J$, and  $G =-iHt/2\hbar +iH\sin (4Jt/\hbar)/8J$. 
Then similar to the above mentioned decimation procedure can be used. It produces
\begin{eqnarray}
&&K' =P +G\left( 1 + {1-e^{-4P}\over 2} \right) \ , \nonumber \\
&&G' = {G\over 4} \left(1-e^{-4P}\right) \ , 
\end{eqnarray}
with $\tanh P = \tanh^2 K$ being the solution for $H=0$.  This system of renormalization group equations has two fixed points $(G^*,K^*)=(0,0)$ and $(G^*,K^*)=(\infty,0)$. In the vicinity of the second (unstable) fixed point one has $\delta K' =2\delta K +3G/2$, and $G' = G/4$. Hence $H \ll J$ is the irrelevant perturbation. Notice that in the exact solution it is necessary to have $H > J$ to destroy the dynamical quantum phase transitions. Also, scaling properties are invariant under a small modification of the initial state by taking the ground state for the initial transverse field $1 < H/J < \infty$. In this case the above mentioned scaling and universality are also valid. 

Now, let us turn to the two-dimensional case. Here one obtains for the density of the Loschmidt amplitude using the exact result for the partition function \cite{Ons,LSM}
\begin{equation}
g(t) = -{1\over 2} \ln [2\sinh K] - {1\over 4\pi} \int_{-\pi}^{\pi} dk s(\varepsilon_k) \varepsilon_k \ , 
\end{equation} 
where
\begin{eqnarray}
&&\cosh \varepsilon_k = \cosh (2K_{\perp} ) \cosh (2{\bar K} ) \nonumber \\
&&- \sinh (2K_{\perp}) \sinh (2{\bar K})\cos k '\ , 
\end{eqnarray}
with $\exp (-3{\bar K}) = \tanh (K)$, and $s(x) ={\rm sign} [{\rm Re} (x)]$. Dynamical quantum phase transitions are found \cite{Hey2} for $J_{\perp} \ll J$ at $t_n^* = \hbar \pi (2n+1)/4J$, i.e., they are controlled by the one-dimensional mechanism. It is clear, because $(J_{\perp}/J) \ll1$ is the irrelevant perturbation. On the other hand, for $J_{\perp} =J$ it follows  
\begin{equation}
g(t) \sim \tau^2 \ln (|\tau|) \ , 
\label{2d}
\end{equation}
i.e., the logarithmic non-analyticity of the dynamical quantum phase transition is revealed. It is, unfortunately, impossible to derive the closed set of renormalization group equations for the two-dimensional case. However, it is possible to perform calculations for the strong spatial anisotropy $J_{\perp} \ll J$. Again, eliminating $K_{\perp}$ perturbatively using the cumulant expansion we get to the second order in $K_{\perp}$ 
\begin{eqnarray}
&&K' = K + 2QK_{\perp}^2 \ , \nonumber \\
&&K'_{\perp} =K_{\perp}^2 \ , 
\end{eqnarray}
where $Q = \tanh K$ for $|\nu_c| > |\nu_s$ and $Q = 1/tanh K$ for the opposite case. If initialy $K_{\perp} <1$, then the fixed point $K^*_{\perp} =0$ is always approached. Then $K_{\perp}$ is the irrelevant perturbation. It implies that in the case of the strong spatial anisotropy the effective dimension is $d^*=1$, i.e., we consider the set of weakly coupled Ising chains. 
For the isotropic point the scaling Eq.~(\ref{2d}) suggests that the unstable fixed point is the governing parameter for dynamical quantum phase transitions. 

Also, the numerical calculations for the dynamics of the order parameter \cite{Hey2} manifest oscillations of the correlation function of the latter. Notice that $H/J \to 0$ is the singular point. The dynamical susceptibility in the limit $H \to 0$  behaves as 
\begin{equation}
\chi(\tau) = \tau^{2d^*} \ . 
\end{equation} 
Here scaling depends only on the effective dimension $d^*$, and only on the universality class of dynamical quantum phase transitions if one assigns $d^*=2$ for the latter for $g(t)$ satisfying the scaling form Eq.~(\ref{2d}). 

Here it is worth to mention the scaling with respect to the connection of the dynamical phase transitions to the universal Kibble-Zurek scaling \cite{KZ} of the defect density and residual energy measured in the final state after the quantum quench \cite{KZq}. It can be studied using the Ising spin-1/2 chain in the tilted magnetic field \cite{SSD}.  When a $d$-dimensional quantum system was initially in its ground state, and then is driving by the quantum quench across the quantum critical point, when the ramping is linear $\sim t/\tau$, then the density of defects  according to the Kibbel-Zurek scaling generalized to quantum critical systems, is proportional to $\tau^{-d \nu/(z\nu +1)}$.  The residual energy (the excess energy over the ground state of the final Hamiltonian) scales as $\varepsilon_{res} \sim \tau^{-(d +z)\nu/(z\nu +1)}$ for the quench to the gapless quantum critical point, and as the density of defects when quenched to the gapped phase \cite{rew}. Using the time-dependent density matrix renormalization group calculations \cite{SSD} obtained for the transverse field Ising chain with the quantum ramping of the component of the field, parallel to the Ising axis, $h$, a very good agreement with the Kibbel-Zurek scaling for the residual energy. Also, according to \cite{rew} the sudden quench for small amplitude of the mentioned parallel field the scaling has to be $\varepsilon_{res} \sim H^{\nu_H (d+z)}$, which has been also confirmed by the numerical calculations \cite{SSD}.

\section{Manifestations of dynamical quantum phase transitions}

Dynamical quantum phase transitions can manifest themselves in the time dependence of several physical characteristics. Out of the ground state the measured values depend not only on the entire ramping protocol, but also on the initial distribution of eigenstates, and, therefore, it is a stochastic variable with some probability distribution. 

However, dynamical quantum phase transitions were not observed in experiments to date, because the Loschmidt echo is not directly related to a quantum mechanical observable. Hence, we need to find some quantum mechanical observables, which time evolution after the quantum quench can manifest features, related to the dynamical quantum phase transitions. 

The work distribution function as a function of time and the performed work $W$ (given by the difference of two consecutive measurements of the energy) is (here we closely follow \cite{HPK, Abel})
\begin{equation}
P(W,\tau) = \sum_n p_{m|n}(\tau)p_n(0)\delta [W-E_m(g_1) -E_n (g_0)] \ , 
\end{equation}
where the energies are measured at $t=0$ and $t=\tau$, $p_n(0)$ is the initial probability of distribution of states, and $p_{n|m}(\tau)$ is the probability to transition from the state $n$ to the state $m$.  In the ground state $p_n(0)=\delta_{n,0}$, i.e., only the ground state ($n=0$) survives. Also, 
\begin{equation}
p_{n|m} = | \langle E_m(g_0) |e^{-i{\cal H}(g_1)t/\hbar}|E_n(g_0)\rangle |^2 \ . 
\end{equation}
For such a distribution function important facts are known, like the Jarzynski equality and the 
Tasaki-Crooks relation \cite{Cam}. Those relations were used to be tested for classical microscopic systems \cite{relexp}, and only recently they were verified for the quantum system of trapped ions \cite{expqu}. The Loschmidt amplitude shows the density of probability that no work is performed on the system during the pulse, since the final and initial states are the same. The sum is over all eigenstates (labelled by $n$) of the initial Hamiltonian. We can introduce the rate function 
\begin{equation}
r(w,\tau) = -(1/L)\ln P(W,\tau) \ , 
\end{equation}
were the work density is $w =W/L$. The rate function is obviously non-negative \cite{GS}. Then, according to G\"artner-Ellis theorem \cite{Tou} the rate function is the Legendre-Fenchel transform $r(w,\tau) = {\rm inf} [c(R,\tau) -wR]$, where the infimum (the greatest lower bound) is evaluated within the considered area, which includes $r=\pm \infty$. For $w=0$ it is just the return probability to the ground state $l(\tau)$. For the transverse field Ising chain one gets 
\begin{eqnarray}
&&c(R,\tau) = -\int_0^{\pi} {dk\over 2\pi} \ln G_k \ , \nonumber \\
&&G_k = [ 1-\sin^2(2\phi_k)\sin^2(\varepsilon_k(H_1)\tau/\hbar)] 
\nonumber \\
&&\times [\exp(-2\varepsilon_k(H_0)R) -1] ] \ . 
\end{eqnarray}
It is simpler than the work distribution function, because it splits into sums over $k$. $c(R,\tau)$ is always concave and continuous inside the relevant considered area. The G\"artner-Ellis theorem states that if $c(R,\tau)$ is differential with respect to $R$, the probability distribution $p(w,\tau)$ also takes on the large deviation form. 

It is nothing else than the rate function for the cumulant generating function of the work distribution 
\begin{equation}
C(R,\tau) = \int dW P(W,\tau) e^{-RW} \equiv e^{-Lc(R,\tau)} \ . 
\end{equation}
This is why, at $R =\infty$ the cumulant generating function has also to manifest Fisher's zeros in the $\tau$-dependence, which signals the dynamical quantum phase transitions for the pulse between two phases (in the sense of the quantum critical point. The generalization for nonzero temperatures is straightforward. However, for $T\ne 0$ the integrand $G_k$ is always analytic for any parameters of quenches (pulses). For the mode $k^*$ at critical values of $\tau =\tau^*_n$ we get (notice that $\phi_{k^*}=\pi/4$) 
\begin{equation}
G_{k^*}(R,\tau_n^*) = {1+\cosh[(-2R +(1/T))\varepsilon_{k^*}(H_1)]\over 1+\cosh (\varepsilon_{k^*}(H_0)/T)} \ .
\end{equation}
Obviously, it is equal to zero only in the zero temperature limit for $R \to \infty$. 

It is possible to check that the Jarzynski relation \cite{Jar}
\begin{equation}
\langle \exp (-W/T) \rangle_{\lambda} = \exp (-\Delta F/T) \ , 
\end{equation}
where the average is taken with respect to initial Gibbs distribution for the ramping protocol $\lambda$, holds. It means that the mean of work values $W$ of many identical experiments is equal to the change of the free energy $\Delta F$ of the corresponding equilibrium states. So we can study the equilibrium states of the system even if it is not in equilibrium during the work measurements. The Tasaki-Crooks relation generalizes the Jarzynski equality as \cite{TC}
\begin{equation}
{p[w,\lambda]\over p[-w,{\tilde \lambda}]} = e^{L (w-\Delta f)/T)} \ , 
\end{equation}
where $\Delta F=L\Delta f$, and $p[w,\lambda]$ denotes the work distribution function of the work density for the ramping protocol $\lambda$. For the rate functions for forward (F) and backward (B) quenches we have $(1/L)\ln [p_F(w)/ p_B(-w)] = r_B(-w) -r_F(w)$. In the pulse case $\Delta F =0$ (i.e., we consider the cyclic process). Hence, $r(w) -r(-w) = -w/T$. Then it is easy to verify the fulfillment of the Jarzynski and Tasaki-Crooks relations directly, i.e, to understand the behavior of the free energy in equilibrium  from the time dependence ($\tau$-dependence) behavior of the $G(z)$ after the quantum quench (pulse). 

In practice more viable route to verify in experiments the phenomenon of the dynamical quantum phase transitions is through measuring the time evolution of some thermodynamic quantities, which can exhibit post-quench oscillations at the time scale commensurate with the critical time $t^*$ of the dynamical quantum phase transitions. 

For example, we can consider the characteristics of the quantum system after the quantum quench, or after the pulse - which is, in fact, two quantum quenches. For the pulse the parameter of a quantum system is initially in one state, then, at time $t=0$ changes its value, analogous to the final state in the quantum quench, and, then, at time $t'=\tau$ returns to the initial value \cite{Zv1,Zv}. For example, the change of the magnetization of the transverse Ising chain at nonzero temperature $T$ after the quantum quench from the state with $H=H_0$ to the state with $H=H_1$ is \cite{Zv1}
\begin{eqnarray}
&&\Delta M^z = {(H_1-H_0)\over 4\pi}\int_0^{\pi} dk \tanh \left({\varepsilon_k(H_0)\over 2T}\right) \times 
\nonumber \\
&&{J^2\sin^2 k \over \varepsilon_k(H_0)\varepsilon^2_k(H_1)} \sin^2(\varepsilon_k(H_1) t/\hbar) \ . 
\label{magn}
\end{eqnarray}
For the pulse case we need to change $t \to \tau$, i.e., the change of the magnetization depends not on time, but on the pulse duration. One can see that Eq.~(\ref{magn}) is similar in structure to Eq.~(\ref{dens1}). Namely, the time dependence is determined by the same oscillatory multiplier. This is why, Fisher's zeros also determine the time evolution of the magnetization of the transverse Ising chain, and the same $t^*$ (related to the imaginary part of Fisher's zeros) defines the periodicity of oscillations. The difference is, naturally, in the absence of kinks in the time dependence ($\tau$-dependence) of the magnetization, especially at nonzero temperature, with respect to the dynamical quantum phase transitions seen in the dynamics of the density of the return probability. On the other hand, the rates of the decays of oscillations of the magnetization and the density of the return probability are determined by different values, which are related to the different multipliers in front of oscillation terms in the integrands in Eq.~(\ref{magn}) and Eq.~(\ref{dens1}). 
\begin{figure}
\begin{center}
\vspace{-15pt}
\includegraphics[scale=0.35]{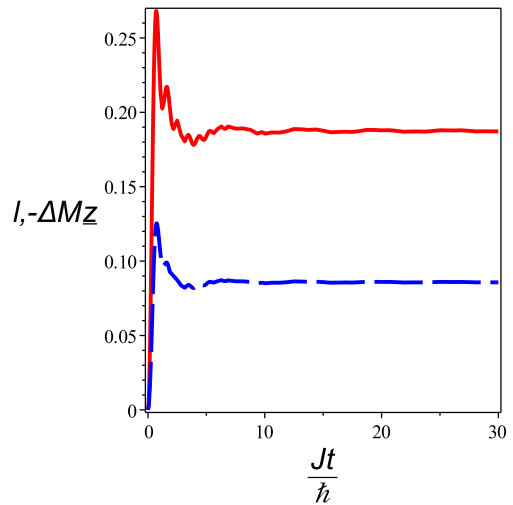}
\end{center}
\vspace{-15pt}
\caption{(Color online) Time dependence of the density of the return probability $l(t)$ (red solid line) and the change of the magnetization $\Delta M^z$ (blue dashed line) of the tranverse field Ising chain (at $T=0.01$) for $J=2$,  
$H_0=0$ and $H_1=-1.5$ (without quantum quench across the quantum critical point $H_c =J$.}   
\label{mis1}
\end{figure}
\begin{figure}
\begin{center}
\vspace{-15pt}
\includegraphics[scale=0.35]{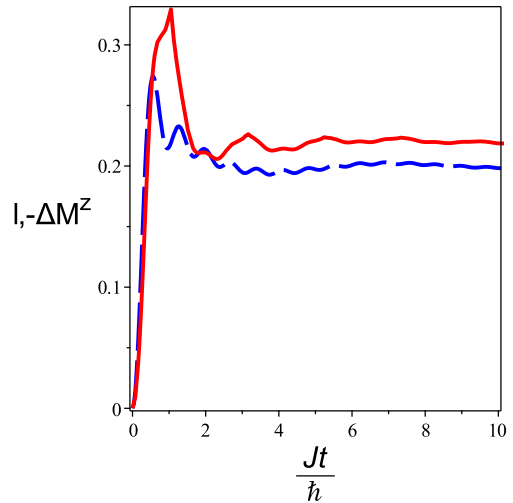}
\end{center}
\vspace{-15pt}
\caption{(Color online) The same as in Fig.~\ref{mis1}, but for $H_1=-2.5$, i.e., the quantum quench across the quantum critical point.}   
\label{mis2}
\end{figure}
\begin{figure}
\begin{center}
\vspace{-15pt}
\includegraphics[scale=0.35]{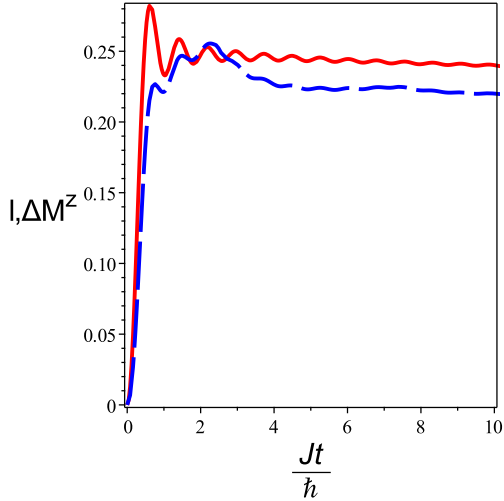}
\end{center}
\vspace{-15pt}
\caption{(Color online) The same as in Fig.~\ref{mis1}, and \ref{mis2} but for $H_1=-2.1$.}   
\label{mis3}
\end{figure}
Figs.~\ref{mis1}-\ref{mis2} show the time dependence of the density of the return probability $l(t)$ and the change of the magnetization $\Delta M^z(t)$ after the quantum quench not across the quantum critical point $H=H_c =J$. There are no kinks in the time dependence of $l(t)$, however we can clearly see the agreement between the periods of oscillations at small time scale (at large time scale the oscillations decay). On the other hand, for the quantum quench across the quantum critical points, the kinks (dynamical quantum phase transitions) are clearly seen in the time dependence of $l(t)$ in Fig.~\ref{mis2}, while the similarity in periodicities of $l(t)$ and $\Delta M^z(t)$ is seen not so clear. Finally, in Fig.~\ref{mis3}, which also corresponds to the quantum quench across the quantum critical point (however, small, comparing to Fig.~\ref{mis2}), the similarity in periodicities of $l(t)$ and $\Delta M^z(t)$ is manifested, while the cusp is seen only for the first dynamical quantum phase transition.  

Similar behavior can be observed in the time dependence of the magnetization of the edge state of the topological insulator and the density of the return probability after the quantum quench across the quantum critical point \cite{Zv}. 
\begin{figure}
\begin{center}
\vspace{-15pt}
\includegraphics[scale=0.35]{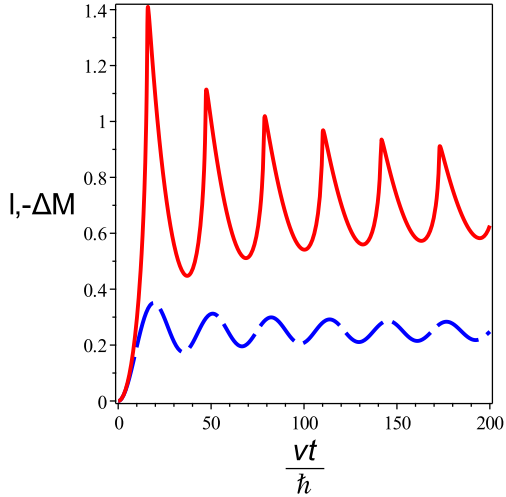}
\end{center}
\vspace{-15pt}
\caption{(Color online) Time dependence of the density of the return probability $l(t)$ (red solid line) and the change of the magnetization $\Delta M$ (blue dashed line) of the one-dimensional edge state of the topological insulator (at $T=0.01$) for the velocity of spin-filtered edge state $v=0.2$,  and 
$H_0=10$ and $H_1=0$ .}   
\label{medge}
\end{figure}
Figs.~\ref{medge} and \ref{medge1} manifest the similarity between the time dependence of the return probability and the change of the magnetization of the edge states of the topological insulator. One can again see clear dynamical quantum phase transitions in the time dependence of $l(t)$ after the quantum quench across the quantum critical point $H=0$, and the agreement in the periodicity of oscillations of the nonzero-temperature magnetization and the density of the return probability (Loschmidt echo). On the other hand, the return probability of the edge state of the topological insulator can show no dynamical quantum phase transitions, even for quantum quenches across the quantum critical point 
\begin{figure}
\begin{center}
\vspace{-15pt}
\includegraphics[scale=0.35]{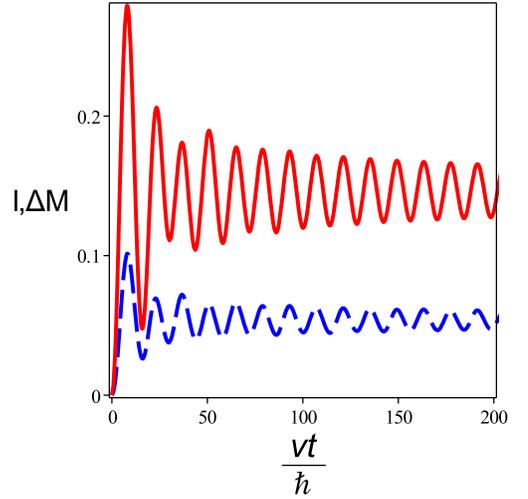}
\end{center}
\vspace{-15pt}
\caption{(Color online) The same as in Fig.~\ref{medge} but for $H_0=0.1$}   
\label{medge1}
\end{figure}
To conclude, for nonzero temperatures the dynamical quantum phase transitions are impossible to observe. However, their traces can be observed in the time (or $\tau$) dependences of either the work distribution function, or the number of particles (magnetization) for quantum quenches ( or pulses) across quantum critical points, as oscillations with periods related to Fisher's zeros.

\section{Summary}

In our review article we have given a brief introduction to the theory of dynamical quantum phase transitions, focusing on the main recent achievements in this rapidly developing part of modern non-equilibrium quantum dynamics. The emphasis is on the universal properties of dynamics, periodicity (related to Fisher'z zeros of the density of the return probability), symmetry aspects, scaling, connection to the topology., etc. We have shown  how zero-temperature dynamical quantum phase transitions can be observed in the experiments on the non-equilibrium time evolution of many-body quantum systems. From this perspective, the most promising candidates for observation of traces of dynamical quantum phase transitions are ultracold atoms in optical traps, and low-dimensional electron systems. In the description of many-body quantum systems out of equilibrium many open questions remain, to which, hopefully, future studies can give clear answers.  

\section{Acknowledgements}

I thank M.~Heyl for interesting discussions related to the nature of the dynamical quantum phase transitions. The support from the DFG via SFB1143 and from the Institute for Chemistry of V.N.~Karazin Kharkov National University is acknowledged.

\end{document}